\documentclass[12pt]{article}
\usepackage{enumitem}
\usepackage{amsmath}
\usepackage{amsthm}
\usepackage{fancyhdr}
\usepackage{mathrsfs}
\usepackage{natbib}
\usepackage[margin=1in]{geometry}
\usepackage{color}
\usepackage{multirow,array}
\usepackage{amsmath,amsthm,amssymb}
{
\theoremstyle{plain}
\newtheorem{theorem}{Theorem}[section] \newtheorem{proposition}{Proposition}[section] \newtheorem{lemma}{Lemma}[section]
\newtheorem{assumption}{Assumption}
\newtheorem{definition}{Definition}

  }
  \usepackage[colorlinks = true,
            linkcolor = blue,
            urlcolor  = blue,
            citecolor = blue,
            anchorcolor = blue]{hyperref}
\usepackage{hyperref}
\usepackage{verbatim}
\usepackage{tikz}
\usepackage{mathtools}

\usepackage{appendix}
\usepackage{float}
\newtheorem{example}{Example}
\usepackage{bbm}
\usepackage{graphicx}
\usepackage{caption}
\usepackage{subcaption}
\usepackage[utf8]{inputenc}
\usepackage[T1]{fontenc}    % use 8-bit T1 fonts
\usepackage{hyperref}       % hyperlinks
\usepackage{url}            % simple URL typesetting
\usepackage{booktabs}       % professional-quality tables
\usepackage{amsfonts}       % blackboard math symbols
\usepackage{nicefrac}       % compact symbols for 1/2, etc.
\usepackage{microtype}      % microtypography
\renewcommand{\baselinestretch}{1.1}

\DeclareMathOperator*{\argmin}{arg\,min}
\let\emptyset\varnothing

\numberwithin{equation}{section}

\newcommand{\cU}{\mathcal U}
\newcommand{\eV}{\mathbf V}
\newcommand{\M}{\mathbb M}
\newcommand{\Mn}{\mathbb{M}_n}

\newcommand{\q}{q}

\newcommand{\RN}[1]{%
  \textup{\uppercase\expandafter{\romannumeral#1}}%
}

\title{Robust Tests of Model Incompleteness in the Presence of Nuisance Parameters\thanks{We thank for their comments Iv{\'a}n Fern{\'a}ndez-Val, Jean--Jacques Forneron, Bryan Graham, Hidehiko Ichimura, Paul Koh, Lixiong Li, Francesca Molinari, Pierre Perron, Zhongjun Qu, Marc Rysman, Katsumi Shimotsu, Kevin Song,  Guang Zhang, and seminar and conference participants at Boston University, Johns Hopkins University, University of Tokyo, and the North American Meeting of the Econometric Society 2023. Kaido gratefully acknowledges financial support through NSF Grant SES-2018498.}}
\author{Shuowen Chen \\ Bates White \\ \url{shuowen.chen@bateswhite.com} 
            \and Hiroaki Kaido \\ Department of Economics \\ Boston University \\ \url{hkaido@bu.edu}}
\begin{document}
\maketitle

\begin{abstract}
Economic models may exhibit incompleteness depending on whether or not they admit certain policy-relevant features such as strategic interaction, self-selection, or state dependence.
We develop a novel test of model incompleteness and analyze its asymptotic properties. A key observation is that one can identify the \emph{least-favorable parametric model} that represents the most challenging scenario for detecting local alternatives without knowledge of the selection mechanism.
We build a robust test of incompleteness on a score function constructed from such a model.
The proposed procedure remains computationally tractable even with nuisance parameters because it suffices to
estimate them only under the null hypothesis of model completeness. We illustrate the test by applying it to a market entry model and a triangular model with a set-valued control function.
\medskip

\end{abstract}

\clearpage

\section{Introduction}
Discrete choice models are used widely. A common empirical strategy is to combine a theory of choice (e.g., utility maximization) that predicts a unique outcome value with distributional assumptions on latent variables \citep{mcfadden1981econometric}. This approach allows the researcher to derive the conditional distribution of the outcome given covariates and apply likelihood-based inference methods. 
However, recent economic applications often involve models that permit multiple outcome values, which we call an \emph{incomplete prediction}.
Such an incomplete prediction occurs when the researcher is willing to work only with weak assumptions or has limited knowledge of the data-generating process. 

This paper considers a form of incompleteness summarized as follows.  An observable discrete outcome $Y\in \mathcal Y$ satisfies
    \begin{align}
        Y\in G(U|X;\theta),\label{eq:inc_model}
    \end{align}
  where $G$ collects all outcome values compatible with the model given observed and unobserved variables $(X,U)$ and a structural parameter $\theta$. This structure arises in a variety of contexts. For example, multiple outcomes are predicted in single-agent discrete choice models when the agent's unobservable choice set is consistent with a wide range of choice set formation processes  \citep{barseghyan2021heterogeneous}. Multiple equilibria may exist in discrete games such as firms' market entry or household labor supply decisions, but one may not know how an equilibrium outcome gets selected \citep{BRESNAHANJoE91,CilibertoTamer2009}.
  Recent empirical studies have  applied inference methods for incomplete models in different areas; they include English auctions \citep{haile2003inference},  strategic voting \citep{kawai2013inferring}, product offerings \citep{eizenberg2014upstream,wollmann2018trucks}, network formation \citep{depaulaEtAl2018,sheng2020}, school choice \citep{fack2019beyond}, and major choice \citep{henry2020revealing}. 
  
This paper focuses on developing tests to determine if a structural model is incomplete.  The completeness of a model is important when considering its policy implications. For example, in a canonical market-entry model, multiple equilibria exist only if firms strategically interact. Testing for strategic interaction effects and inferring their signs can provide valuable information for policymakers \citep{depaulatang2012}. In a triangular model with a discrete endogenous variable, the control function approach yields an incomplete model, but it remains complete if the assignments are exogenous. Detecting self-selection can aid practitioners in selecting a proper strategy for evaluating treatment effects.  Dynamic discrete choice models can make incomplete predictions when they admit state dependence, but they are complete without it. 
The presence of state dependence can significantly impact program evaluation and counterfactual analyses \citep{card2005estimating,handel2013adverse}.

 In many of these examples, one can state the null hypothesis of model completeness as a restriction on the parameter's subvector. We develop a novel score test for such a restriction.
 Advantages of this approach are (i) the score statistic only requires estimation of nuisance parameters in the complete restricted model; (ii) one can use package software to estimate the nuisance parameters; and (iii) 
simulating the score statistic's limiting distribution is straightforward.

Our focus is on models that are complete under the null hypothesis. This structure makes score-based tests appealing. First, one can obtain a unique likelihood function $q_{\theta_0}$ for any null parameter value $\theta_0$, thanks to the model's completeness. For each alternative parameter value $\theta_0+h$, the model implies multiple (typically infinitely-many) likelihood functions. However, the recent work of \cite{kz} demonstrates that one can identify a "least favorable" density $q_{\theta_0+h}$ that is most difficult to distinguish from $q_{\theta_0}$. We may then consider the family $\{q_{\theta_0+h}\}_{h\in\mathbb R^d}$ as a "least favorable parametric model". Second, we can use the least favorable parametric family to calculate a score function. We then construct a score-based statistic that maximizes a measure of local discrimination. The resulting test is robust to incompleteness because it detects any local deviation from the null hypothesis regardless of how $Y$ is selected from the predicted set.
 
The score test has another appealing feature wherein one can estimate nuisance parameters within the complete restricted model. Typically, the null hypothesis does not restrict some components of $\theta$. By exploiting the model completeness, we demonstrate that one can construct a restricted maximum likelihood estimator (RMLE) of these nuisance components that is $\sqrt n$-consistent. This estimator is usually simple to calculate using package software. Next, we insert the estimator into the score formula to compute a test statistic. The suggested procedure is computationally tractable since it avoids evaluating the test statistic over a grid of nuisance parameters. Finally, we derive the score-based statistic's limiting distribution and demonstrate that it can be easily simulated.
Although there are existing methods to test restrictions on subvectors of $\theta$, they can be computationally expensive since they are not designed to utilize the model completeness under the null hypothesis. This paper presents a procedure that makes use of the model structure to simplify its implementation. Through a Monte Carlo experiment, we also show that the proposed test has significantly higher power than a general subvector test that does not utilize the structure.

\subsection{Relation to the Literature}
Our paper belongs to the literature on inference in incomplete models pioneered by \cite{Wald1950} in the context of simultaneous equations models and by \cite{jovanovich89} in the context of models with multiple equilibria. The seminal work of \cite{tamer} shows
an incomplete model induces multiple distributions and implies partially identifying restrictions on parameters. Developments in the literature \citep{galichon2011set,bmm,chesher2017generalized} provided tools to systematically derive so-called \emph{sharp identifying restrictions}, which convert all model information into a set of equality and inequality restrictions on the conditional moments of observable variables. Inference methods based on the sample analogs of moment restrictions are extensively studied \citep[see][and references therein]{canay/shaikh:2017}. Our approach builds on recent developments in likelihood-based inference methods for incomplete models \citep{Chen2018,kz}.
In particular, we combine the sharp identifying restrictions with the framework of \cite{kz} to derive the least favorable parametric model and its score.
To our knowledge, this approach is new. One can view our procedure as an analog of deriving a score function from a parametric family in a complete model.

Hypothesis testing in incomplete models is studied extensively. 
As discussed earlier, many of them are based on the sample analogs of conditional or unconditional moment restrictions. A challenge in making inferences is the high computational cost of implementing existing methods, as noted in \cite{molinari2020microeconometrics}. There are attempts to improve the computational tractability of moment-based inference methods within specific classes of models or testing problems, such as those made by \cite{ARP} and \cite{Cox2020}, who assume that moment inequality restrictions are linear conditional on observable variables.  This paper focuses on another class in which the model is complete under the null hypothesis. This structure makes our test computationally tractable by combining (i) the score function associated with the least favorable parametric model and (ii)  a point estimator of the nuisance components.

Practitioners can use this paper's framework to test various hypotheses. For example, one can examine the exsitence of strategic interaction effects and multiple equilibria in static complete information games. Related problems are studied in other models.
For incomplete information games, \cite{depaulatang2012} introduced a semiparametric inference procedure on the signs of strategic interaction effects. For finite-state Markov games, \cite{OtsuPT2016} developed techniques to test whether the conditional choice probabilities, state transition, and other features of games are homogeneous across cross-sectional units. Rejecting their null hypothesis may indicate the presence of multiple equilibria.
\cite{pelican2020optimal}  studied a testing problem that involves determining whether agents' preferences are interdependent in a network formation model. The problem includes nuisance parameters that account for degree heterogeneity and homophily. Using the logit structure, they constructed a sufficient statistic for the nuisance parameters and developed a conditional score test. This paper and  \cite{pelican2020optimal} consider testing in settings with a complete model under the null and an incomplete model under the alternative. The two papers take different approaches by exploiting the structures of the respective models.  Specifically, we use the least favorable parametric model and estimate nuisance parameters using the restricted MLE. In contrast,  \cite{pelican2020optimal} utilized a sufficient statistic for the nuisance parameters.

 One can apply our framework to triangular systems involving a binary outcome and a discrete endogenous variable. We show that taking a control function approach in such a setting leads to a model with an incomplete prediction.\footnote{A nonparametric identification analysis based on set-valued control functions is undertaken in another paper.} 
We provide a test of endogenous treatment assignments under weak assumptions. To our knowledge, this test is new and provides an alternative to the existing proposal by \cite{wooldridge2014JoE}, who makes additional high-level assumptions.

\section{Set-up}
Let $Y$ be a discrete outcome taking values in a finite set $\mathcal Y$.  Let $X\in\mathcal X\subseteq\mathbb R^{d_X}$ be a vector of observable covariates and let $U\in \cU\subseteq\mathbb R^{d_U}$ be a vector of unobservable variables. We equip $\mathcal Y,\mathcal X$, and $\cU$ with their Borel $\sigma$-algebra. In what follows, we use upper case letters for random elements (e.g., $X$) and lower case letters (e.g., $x$) for the values they can take. 
Let  $\theta\in\Theta\subset\mathbb R^{d_\theta}$ be a finite-dimensional parameter, where $\Theta$ is a convex parameter space with a nonempty interior.

A set-valued map $G:\cU\times\mathcal X\times\Theta\leadsto \mathcal Y$ summarizes the prediction of a structural model.
We assume $G(\cdot|\cdot;\theta)$ is weakly measurable for every $\theta\in\Theta$ and $Y$ takes one of the values in $G(U|X;\theta)$ with probability 1.\footnote{A set-valued map is weakly measurable if its weak inverse image $G_{-1}(A)\equiv\{s\in\mathcal S :G(s)\cap A\ne\emptyset\}$ is measurable for any open set $A$.} A random element $Y$ with this property is said to be a \emph{measurable selection} of the random closed set $G(U|X;\theta)$ \citep[][]{molchanov2005theory}.
The map $G$ describes how observable and unobservable characteristics translate into a set of possible outcome values. It reflects restrictions imposed by theory, such as the functional form of utility/profit functions, forms of strategic interaction, and any equilibrium or optimality concepts. Importantly, $G(u|x,\theta)$ can contain multiple values.
This feature allows the researchers to encode their lack of understanding of parts of the structural model.

The formulation above also nests the standard setting in which the model is characterized by a \emph{reduced form equation}:
\begin{align}
    Y=g(U|X;\theta),
\end{align}
for a function $g:\cU\times \mathcal X\times\Theta\to \mathcal Y$. In this case,  $G$ is almost surely singleton-valued, i.e., $G(U|X;\theta)=\{g(U|X;\theta)\}$, and we say the model makes a \emph{complete prediction}.

Throughout, we assume $U$'s law belongs to a parametric family $F=\{F_\theta,\theta\in\Theta\}$, where, for each $\theta$, $F_\theta$ is a probability distribution on $\cU$. To keep notation concise, we use the same $\theta$ for parameters that enter $G$ and that index $F_\theta$. Also, we focus on settings in which $U$ is independent of $X$. However, the framework can be easily extended to settings where $U$ is correlated with $X$, and the researcher specifies its conditional distribution $F_\theta(u|x)$. Furthermore, our framework accommodates settings in which some of the observable covariates are endogenous, and one can construct a set-valued control function (See Example \ref{ex:incomplete_cf} below).

\subsection{Motivating examples}\label{sec:examples}
We illustrate the objects introduced above with examples. Our first example is a  discrete game of complete information \citep{BRESNAHANJoE91,CilibertoTamer2009}.

\noindent

\begin{figure}[h]
	\centering
	\caption{Level sets of $u\mapsto G(u|x;\theta)$}
	\begin{tikzpicture}
	[scale=0.8,domain=-3:3,>=latex] 
	\fill[fill=red,opacity=0.4] (-3,-1) -- (-1,-1) -- (-1,3) -- (-3,3) ;
	\fill[fill=green,opacity=0.4] (-1,-3) -- (-1,-1) -- (3,-1) -- (3,-3) ;
	\draw[->] (-3,0) -- (3,0) node[right] {$u_1$}; 
	\draw[->] (0,-3) -- (0,3) node[above] {$u_2$}; 
	\draw[thick, dashed] (-1,-3) -- (-1,3);
	\draw[thick, dashed] (-3,-1) -- (3,-1);
	\draw (-1,-0.5) node[right]{$A$} ; 
	\draw (1,2) node[right] {\small $\{(1,1)\}$}; 
	\draw (1,-2) node[right] {\small $\{(1,0)\}$}; 
	\draw (-1,2) node[left] {\small $\{(0,1)\}$}; 
	\draw (-1,-2) node[left] {\small $\{(0,0)\}$}; 
	\filldraw (-1,-1) circle (2pt);
	\end{tikzpicture}
	\hspace{0.2in}
	\begin{tikzpicture}
	[scale=0.8, domain=-3:3,>=latex]
	\draw[->] (-3,0) -- (3,0) node[right] {$u_1$}; 
	\draw[->] (0,-3) -- (0,3) node[above] {$u_2$}; 
	\draw[thick, dashed] (-1,-3) -- (-1,1.1);
	\draw[thick, dashed] (-3,-1) -- (1.2,-1);
	\filldraw (-1,-1) circle (2pt);
	\draw (-1,-0.5) node[right]{$A$} ; 
	\draw[thick, dashed] (1.2,-1) -- (1.2,3);
	\draw[thick, dashed] (-1,1.1) -- (3,1.1);
	\filldraw (1.2,1.1) circle (2pt);
	\draw (1.25,1.4) node[right]{$B$}; 
	\fill[fill=red,opacity=0.4] (1.2,3) -- (-3,3) -- (-3,1.1) -- (1.2,1.1);
	\fill[fill=red,opacity=0.4] (-1,1.1) -- (-3,1.1) -- (-3,-1) -- (-1,-1);
	\fill[fill=green,opacity=0.4] (-1,-1) -- (-1,-3) -- (3,-3) -- (3,-1);
	\fill[fill=green,opacity=0.4] (1.2,-1) -- (3,-1) -- (3,1.1) -- (1.2,1.1);
	\draw (-1,-2) node[left] {\small $\{(0,0)\}$};  
	\draw (1.3,2) node [right]{\small $\{(1,1)\}$};
	\draw (-1,2) node[left] {\small $\{(0,1)\}$}; 
	\draw (1.3,-2) node[right] {\small $\{(1,0)\}$}; 
	\draw (-0.8,0.2) node [right]{\small $\{(1,0),$}; 
	\draw (-0.2,-0.5) node [right]{\small $(0,1)\}$}; 
	\fill[fill=blue,opacity=0.4] (-1,-1) -- (1.2,-1) -- (1.2,1.1) -- (-1,1.1);
	\end{tikzpicture}
	\begin{minipage}{0.8\textwidth}
		
	\bigskip
	{\footnotesize Note: $A=(-x^{(1)}{}{'}\delta^{(1)}, -x^{(2)}{}{'}\delta^{(2)})$; $B=(-x^{(1)}{}{'}\delta^{(1)}-\beta^{(1)}, -x^{(2)}{}{'}\delta^{(2)}-\beta^{(2)})$.
	
	Left Panel: $\beta^{(1)}=\beta^{(2)}=0$ and the model is complete. Right panel: $\beta^{(1)}<0$ and $\beta^{(2)}<0$ and the model is incomplete. $S_{\theta,(1,0)}$ in \eqref{eq:strategic_sub} corresponds to the region in green, and similarly $S_{\theta,(0,1)}$ is the region in red. Multiple equilibria $\{(1,0),(0,1)\}$ are predicted in the blue region.  \par}
	\end{minipage}
	\label{fig:example1}
\end{figure}
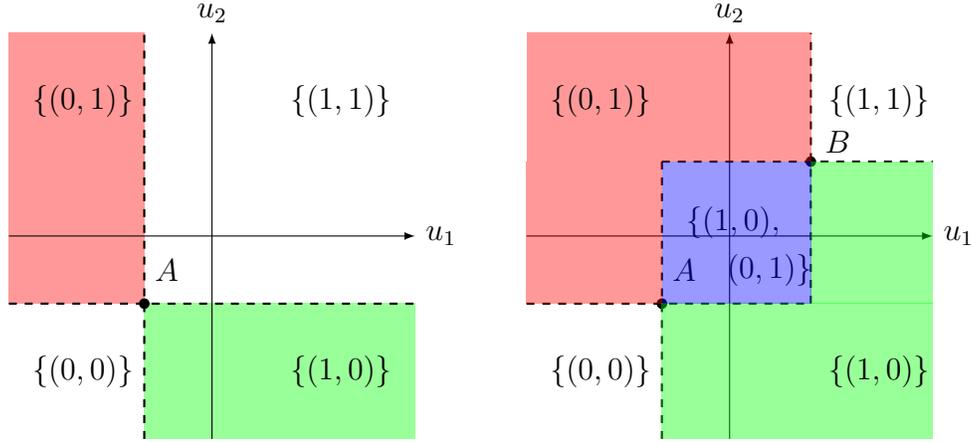

\begin{example}[Discrete Games of Strategic Substitution]\label{ex:ci_game}\rm
There are two players (e.g., firms). Each player may either choose $y^{(j)}=0$ or $y^{(j)}=1$. The payoff of player $j$ is 
\begin{align}
\pi^{(j)}=y^{(j)}\big(x^{(j)}{}{'}\delta^{(j)}+\beta^{(j)}y^{(-j)}+u^{(j)}\big),\label{eq:payoff}
\end{align}
where $y^{(-j)}\in\{0,1\}$ is the opponent's action, $x^{(j)}$ is player $j$'s observable characteristics, and $u^{(j)}$ is an unobservable payoff shifter. The payoff is summarized in the table below and is assumed to belong to the players' common knowledge.
\begin{table}[H]
    \centering
    \setlength{\extrarowheight}{2pt}
    \begin{tabular}{cc|c|c|}
        & \multicolumn{1}{c}{} & \multicolumn{2}{c}{Player $2$}\\
        & \multicolumn{1}{c}{} & \multicolumn{1}{c}{$y^{(2)}=0$}  & \multicolumn{1}{c}{$y^{(2)}=1$} \\\cline{3-4}
        \multirow{2}*{Player $1$}  & $y^{(1)}=0$ & $0, 0$ & $0, x^{(2)}{}{'}\delta^{(2)}+u^{(2)}$ \\\cline{3-4}
        & $y^{(1)}=1$ & $x^{(1)}{}{'}\delta^{(1)}+u^{(1)}, 0$ & $x^{(1)}{}{'}\delta^{(1)}+\beta^{(1)}+u^{(1)}$, $x^{(2)}{}{'}\delta^{(2)}+\beta^{(2)}+u^{(2)}$ \\\cline{3-4}
    \end{tabular}
\end{table}
The key parameter is the \emph{strategic interaction effect} $\beta^{(j)}$ which captures the impact of the opponent's taking $y^{(-j)}=1$ on player $j$'s payoff.
Suppose that $\beta^{(j)}\le 0$ for both players.\footnote{This restriction is often used in models of market entry. Games with strategic complementarity (i.e., $\beta^{(j)}\ge 0$) can be analyzed similarly.} 
Suppose that the players play a pure strategy Nash equilibrium (PSNE). Then, one can summarize the set of PSNEs by the following correspondence:
\begin{align}
G(u|x;\theta)=
\begin{cases}
\{(0, 0)\} & u^{(1)}<-x^{(1)}{}'\delta^{(1)}, u^{(2)}<-x^{(2)}{}'\delta^{(2)},\\
\{(0, 1)\} & u\in S_{\theta,(0,1)},\\
\{(1, 0)\} & u\in S_{\theta,(1,0)},\\
\{(1, 1)\} & u^{(1)}>-x^{(1)}{}'\delta^{(1)}-\beta^{(1)}, u^{(2)}>-x^{(2)}{}'\delta^{(2)}-\beta^{(2)},\\
\{(1, 0), (0, 1)\} & -x^{(j)}{}'\delta^{(j)}<u^{(j)}<-x^{(j)}{}'\delta^{(j)}-\beta^{(j)}, \quad j=1, 2,
\end{cases}\label{eq:strategic_sub}
\end{align}
where, $\theta = (\beta', \delta')'$,  $S_{\theta,(1,0)}=\{u^{(1)}>-x^{(1)}{}'\delta^{(1)}-\beta^{(1)}, u^{(2)}<-x^{(2)}{}'\delta^{(2)}-\beta^{(2)}\}\cup\{-x^{(1)}{}'\delta^{(1)}<u^{(1)}<-x^{(1)}{}'\delta^{(1)}-\beta^{(1)}, u^{(2)}<-x^{(2)}{}'\delta^{(2)}\}$ and $S_{\theta,(0,1)}=\{u^{(1)}<-x^{(1)}{}'\delta^{(1)}, u^{(2)}>-x^{(2)}{}'\delta^{(2)}\}\cup\{-x^{(1)}{}'\delta^{(1)}<u^{(1)}<-x^{(1)}{}'\delta^{(1)}-\beta^{(1)}, u^{(2)}>x^{(2)}{}'\delta^{(2)}-\beta^{(2)}\}$. 

Figure \ref{fig:example1} shows the $U$-level sets of  $G$ for a given $(x,\theta)$.  
 When $\beta^{(j)}=0$ for either of the players, the model predicts a unique equilibrium for any value of $u=(u^{(1)},u^{(2})'$ (left panel of Figure \ref{fig:example1}).
When $\beta^{(j)}<0$ for both players, the model predicts  multiple equilibria $\{(0,1),(1,0)\}$ when each $u^{(j)}$ is between the two thresholds $x^{(j)}{}'\delta^{(j)}$ and $x^{(j)}{}'\delta^{(j)}-\beta^{(j)}$ (the blue region in Figure \ref{fig:example1}).
\end{example}

The next example is a parametric version of triangular nonseparable equations with  binary outcome and treatment \citep{chesher2003,shaikh_vytlacil2011}.  We consider a \emph{control function approach} applied to the triangular system.

\begin{example}[Triangular Models with a Set-valued Control Function]\label{ex:incomplete_cf}\rm
Consider a triangular model, in which a binary outcome  $Y_{i}$ is determined by a binary treatment  $D_i$, a vector $W_i$ of exogenous covariates, and an unobserved variable $\epsilon_i$. The binary treatment  $D_i$ is determined by a vector of instrumental variables $Z_i$ and an unobserved variable $V_i$.  
\begin{align}
    Y_{i}&=1\{\alpha D_i+W_i'\eta+ \epsilon_i\ge 0 \}\label{eq:outcome},\\
    D_i&=1\{Z_i'\gamma+V_i\ge0\}.\label{eq:inc_cf1}
\end{align} 
The unobserved variables $(\epsilon_i,V_i)$ may be dependent rendering $D_i$ potentially endogenous. We assume that $(W_i,Z_i)$ is independent of $(\epsilon_i,V_i)$.

If one could recover  $V_i$ from the observables (which would be possible with a continuous $D_i$), conditioning on $V_i$ would make $\epsilon_i$  independent of $D_i$, i.e. $\epsilon_i|D_i,V_i\sim \epsilon_i|V_i$. This \emph{control function approach} would allow us to recover structural parameters \citep{Blundell:2004td,imbens2009identification,wooldridgeJHR}. With a binary endogenous variable, we cannot uniquely recover $V_i$.\footnote{Alternatively, \cite{wooldridge2014JoE} uses the generalized residual $r_i=d_i\lambda(z_i'\gamma)-(1-d_i)\lambda(-z_i'\gamma)$ from the first stage MLE, where $\lambda$ is the inverse Mills ratio. He makes additional high-level assumptions so that $r_i$ is a sufficient statistic for capturing the endogeneity of $d_i$ and proposes an estimator of the average structural function. Instead of taking this approach, we explore what can be learned from the set-valued control function.} However, the model restricts $V_i$ to the following \emph{set-valued control function}:
\begin{align}
     \eV(D_i,Z_i;\gamma)
    &\equiv\begin{cases}
    [-Z_i'\gamma,\infty)&\text{ if }D_i=1\\
    (-\infty,-Z_i'\gamma]&\text{ if }D_i=0.
    \end{cases}
\end{align}
This set contains the actual control function $V_i$ as its measurable selection.
Suppose that the conditional distribution $\epsilon_i|V_i$ belongs to a location family, in which the location parameter is $\beta V_i$. Then, one may write $\epsilon_i=\beta V_i+U_i$ for some $U_i$ independent of $(D_i,V_i)$.
Substituting this expression into \eqref{eq:outcome}, we can summarize the model's prediction by
\begin{align}
    G(u_i|x_i;\theta)=\Big\{y_i\in\{0,1\}:y_{i}=1\{\alpha d_i+w_i'\eta+ \beta v_i+u_i\ge 0 \},\text{ for some } v_i\in \eV(d_i,z_i;\gamma)\Big\},
\end{align}
where $x_i=(d_i,w_i',z_i')'$ and $\theta=(\beta,\delta')'$ with $\delta=(\alpha,\eta',\gamma')'$. When $\beta$ is positive, one can simplify $G$ further:
\begin{align}
    G(u_i|1,w_i,z_i;\theta)=\begin{cases}
    \{1\} & u_i\ge -\alpha d_i-w_i'\eta+\beta z_i'\gamma\\
    \{0,1\} & u_i< -\alpha d_i-w_i'\eta+\beta z_i'\gamma.
    \end{cases}
\end{align}
As we show below, the control function approach allows one to test the endogeneity of $D_i$ even if the control function is set-valued.\footnote{We take a control function approach that conditions on $v_i$, which only requires specification of the conditional distribution of $\epsilon_i$ given $v_i$. Alternatively, one could take the vector $(y_i,d_i)$ as endogenous variables and specify the joint distribution of $(\epsilon_i,v_i)$. This alternative approach with a stronger assumption would imply a complete model \citep{lewbel2007ier}.}
\end{example}

\bigskip
The next example is a panel dynamic discrete choice model \citep{heckman78,chamberlain_1985,hyslop99}. 
\begin{example}[Panel Dynamic Discrete Choice Models]\label{ex:paneldc}\rm
An individual makes binary decisions across multiple periods according to
\begin{align}
    Y_{it}=1\{X_{it}'\eta+Y_{it-1}\beta+\alpha_i+\epsilon_{it}\ge 0\},~i=1,\dots, n, ~t=1,\dots,T,\label{eq:panel_ddc1}
\end{align}
where $Y_{it}$ is a binary outcome for individual $i$ in period $t$, $X_{it}$ is a vector of observable covariates, $\alpha_i$ is an unobservable individual specific effect, and $\epsilon_{it}$ is an unobserved idiosyncratic error. We use $a_i\in\mathbb R$ and $e_{it}\in\mathbb R$ to denote realizations of $\alpha_i$ and $\epsilon_{it}$ respectively.
If $\beta$ is nonzero, the individual's choice in period $t$ depends on her past choice, rendering the decision \emph{state dependent}. 

 Suppose the researcher observes $(Y_{it},X_{it})$ for $i=1,\dots,n$ and $t=1,\dots,T$.  
Without any knowledge of $Y_{i0}$, the dynamic restrictions \eqref{eq:panel_ddc1} alone do not fully determine the value of $Y_{i1},\dots,Y_{iT}$ \citep{heckman78,heckman1987incidental,honore2006bounds}.\footnote{As an alternative, one could work with the likelihood function conditional on the initial observation. However, this approach can be problematic if one wants to be internally consistent across different periods \citep{honore2006bounds,wooldridge2005simple}.} 
Consider $T=2$. Suppose for the moment $y_{i0}=0$. For a given $(x_i,a_i, e_{i1},e_{i2})$ and $(\beta,\eta)$, the outcome $y_i=(y_{i1},y_{i2})$ must satisfy
\begin{align}
    % y_{i0}&=0\label{eq:panel1}\\
    y_{i1}&=1\{x_{i1}'\eta+a_i+e_{i1}\ge 0\}\label{eq:panel2}\\
    y_{i2}&=1\{x_{i2}'\eta+y_{i1}\beta+a_i+e_{i2}\ge 0\}.\label{eq:panel3}
\end{align}
Similarly, if $y_{i0}=1$, the outcome must satisfy
\begin{align}
    % y_{i0}&=1\label{eq:panel4}\\
    y_{i1}&=1\{x_{i1}'\eta+\beta+a_i+e_{i1}\ge 0\}\label{eq:panel5}\\
    y_{i2}&=1\{x_{i2}'\eta+y_{i1}\beta+a_i+e_{i2}\ge 0\}.\label{eq:panel6}
\end{align}
Without further assumptions, the model permits both possibilities.
Let $U_i=(U_{i1},U_{i2})'$ with $U_{it}=\alpha_i+\epsilon_{it}$. One can summarize the model prediction by
\begin{align}
    G(u_i|x_i;\theta)=\Big\{y_i=(y_{i1},y_{i2})\in \{0,1\}^2: y_i\text{ satisfies either \eqref{eq:panel2}-\eqref{eq:panel3} or \eqref{eq:panel5}-\eqref{eq:panel6}}\Big\}.
\end{align}
If $\beta\ge 0$, one can express this correspondence  as follows:\footnote{Appendix \ref{sec:panel_ddc} provides details and a graphical illustration of $G$. }
\begin{align}
        G(u_i|x_i;\theta)=\begin{cases}
    \{(0,0)\} &  u_{i1}<-x_{i1}'\eta-\beta,~u_{i2}<-x_{i2}'\eta,\\
    \{(0,1)\} &  u_{i1}<-x_{i1}'\eta-\beta,~u_{i2}\ge-x_{i2}'\eta,\\
    \{(1,0)\} &  u_{i1}\ge-x_{i1}'\eta,~u_{i2}<-x_{i2}'\eta-\beta,\\
    \{(1,1)\} &  u_{i1}\ge-x_{i1}'\eta,~u_{i2}\ge-x_{i2}'\eta-\beta,\\
    \{(0,0),(1,0)\} & -x_{i1}'\eta-\beta\le u_{i1}< -x_{i1}'\eta,~ u_{i2}\le -x_{i2}'\eta-\beta,\\
    \{(0,0),(1,1)\} & -x_{i1}'\eta-\beta\le u_{i1}< -x_{i1}'\eta,~  -x_{i2}'\eta-\beta\le u_{i2}<-x_{i2}'\eta,\\
    \{(0,1),(1,1)\} & -x_{i1}'\eta-\beta\le u_{i1}< -x_{i1}'\eta,~ u_{i2}\ge-x_{i2}'\eta.
    \end{cases}
\end{align}
Figure \ref{fig:ddc_levelset} summarizes $G$.
The model makes a complete prediction when there is no state dependence, i.e., $\beta=0$ (left panel).

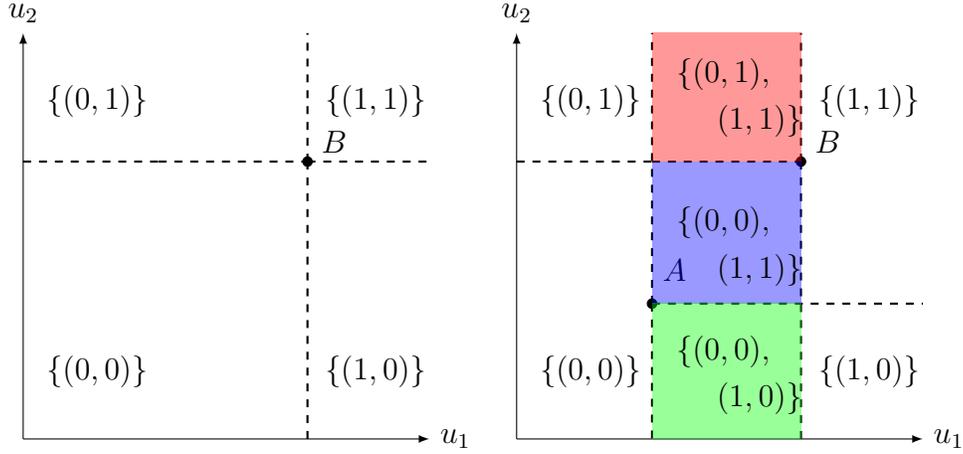
\begin{figure}[h]
	\centering
	\caption{Level sets of $u\mapsto G(u|x;\theta)$ when $\beta\ge 0$}
	\begin{tikzpicture}
	[scale=0.8, domain=-3:3,>=latex]
	\draw[->] (-3,-3) -- (3,-3) node[right] {$u_1$}; 
	\draw[->] (-3,-3) -- (-3,3) node[above] {$u_2$}; 
	\draw[thick, dashed] (-3,1.1) -- (-1,1.1);
% 	\draw[thick, dashed] (-1,-1) -- (-1,3);
 	\draw[thick, dashed] (1.2,1.1) -- (3,1.1);
% 	\filldraw (-1,-1) circle (2pt);
% 	\draw (-1,-0.5) node[right]{$A$} ; 
	\draw[thick, dashed] (1.2,-3) -- (1.2,1.1);
	\draw[thick, dashed] (-1,1.1) -- (1.2,1.1);
	\draw[thick, dashed] (1.2,1.1) -- (1.2,3);
% 	\draw[thick, dashed] (-1,-1) -- (-1,-3);
	\filldraw (1.2,1.1) circle (2pt);
	\draw (1.25,1.4) node[right]{$B$}; 
	\draw (-1,-2) node[left] {\small $\{(0,0)\}$};  
	\draw (1.3,2) node [right]{\small $\{(1,1)\}$};
	\draw (-1,2) node[left] {\small $\{(0,1)\}$}; 
	\draw (1.3,-2) node[right] {\small $\{(1,0)\}$}; 
	\end{tikzpicture}
	\begin{tikzpicture}
	[scale=0.8, domain=-3:3,>=latex]
	\draw[->] (-3,-3) -- (3,-3) node[right] {$u_1$}; 
	\draw[->] (-3,-3) -- (-3,3) node[above] {$u_2$}; 
	\draw[thick, dashed] (-3,1.1) -- (-1,1.1);
	\draw[thick, dashed] (-1,-1) -- (-1,3);
	\draw[thick, dashed] (-1,-1) -- (3,-1);
	\filldraw (-1,-1) circle (2pt);
	\draw (-1,-0.5) node[right]{$A$} ; 
	\draw[thick, dashed] (1.2,-3) -- (1.2,1.1);
	\draw[thick, dashed] (-1,1.1) -- (1.2,1.1);
	\draw[thick, dashed] (1.2,1.1) -- (1.2,3);
	\draw[thick, dashed] (-1,-1) -- (-1,-3);
	\filldraw (1.2,1.1) circle (2pt);
	\draw (1.25,1.4) node[right]{$B$}; 
	\fill[fill=blue,opacity=0.4] (-1,-1) -- (1.2,-1) -- (1.2,1.1) -- (-1,1.1);
	\fill[fill=red,opacity=0.4] (-1,1.1) -- (1.2,1.1) -- (1.2,3) -- (-1,3);
	\fill[fill=green,opacity=0.4] (-1,-1) -- (1.2,-1) -- (1.2,-3) -- (-1,-3);
	\draw (-1,-2) node[left] {\small $\{(0,0)\}$};  
	\draw (1.3,2) node [right]{\small $\{(1,1)\}$};
	\draw (-1,2) node[left] {\small $\{(0,1)\}$}; 
	\draw (1.3,-2) node[right] {\small $\{(1,0)\}$}; 
	\draw (-0.8,0.2) node [right]{\small $\{(0,0),$}; 
	\draw (-0.2,-0.5) node [right]{\small $(1,1)\}$}; 
	\draw (-0.8,2.4) node [right]{\small $\{(0,1),$}; 
	\draw (-0.2,1.7) node [right]{\small $(1,1)\}$}; 
	\draw (-0.8,-1.7) node [right]{\small $\{(0,0),$}; 
	\draw (-0.2,-2.4) node [right]{\small $(1,0)\}$}; 
	\end{tikzpicture}
	\begin{minipage}{0.7\textwidth}
		
	\bigskip
	{\footnotesize Note: The level sets of $G$ when $\beta=0$ (left) and $\beta>0$ (right). $A=(-x_{i1}{}{'}\eta-\beta, -x_{i2}{}{'}\eta-\beta)$; $B=(-x_{i1}{}{'}\eta, -x_{i2}{}{'}\eta)$.
	
    Multiple outcome values are predicted in the red, blue, and green regions.  \par}
	\end{minipage}
	\label{fig:ddc_levelset}
\end{figure}

\end{example}

\section{Testing Hypotheses}\label{sec:Testing}
Let $\beta\in\Theta_\beta\subset\mathbb R^{d_\beta}$ denote the subvector of $\theta$ whose value determines whether the model is complete or not. Let $\delta\in\Theta_\delta\subset\mathbb R^{d_\delta}$ collect the remaining components of $\theta$. Given a sample of data $(Y_i,X_i),i=1,\dots, n$,
consider testing 
\begin{align}
H_0:\beta=\beta_0,~~\text{v.s.}~~H_1:\beta\in B_1,\label{eq:hypotheses}
\end{align}
where $B_1\subset\Theta_\beta$ is a set not containing $\beta_0$. 
For instance, in entry games (i.e. Example \ref{ex:ci_game}), the presence of strategic substitution effects can be tested by letting $\beta_0=0$ and $B_1=\{\beta:\beta^{(j)}<0,j=1,2\}$.\footnote{Our framework also nests settings in which the researcher tests one of the interaction effects, e.g., $\beta^{(1)}_0=0$ and $B_1=\{\beta:\beta^{(1)}<0\}$. In this case, the model is complete under both hypotheses. Our test then reduces to a conventional score test.} Similarly, we may test the potential endogeneity of treatment assignments (Example \ref{ex:incomplete_cf}) and the presence of state dependence (Example \ref{ex:paneldc}) by setting $\beta_0=0$ and choosing suitable alternative hypotheses.
In what follows, we let $\Theta_0=\{\beta_0\}\times \Theta_\delta$ and  $\Theta_1=B_1\times\Theta_\delta$.

Let $\Delta_{Y|X}$ denote the set of conditional distributions of $Y$ given $X$.  For each $\theta=(\beta',\delta')'$, an incomplete model admits the following  set of conditional distributions:
\begin{multline}
	\mathcal Q_\theta=\Big\{Q\in \Delta_{Y|X}:Q(A|x)=\int_U \eta(A|x,u)dF_\theta(u),~\forall A\subseteq \mathcal Y,\\
	~\text{for some }\eta\in\Delta_{Y|X,U} ~\text{such that }\eta(G(u|x;\theta)|x,u)=1,~ a.s.\Big\}.
\end{multline}
The conditional distribution $\eta(\cdot|x,u)$ represents the unknown \emph{selection mechanism} according to which an outcome gets selected from $G(u|x;\theta)$. Reflecting the lack of understanding of the selection,  we allow any law supported on  $G(u|x;\theta)$.
Consequently, the model can admit (infinitely) many likelihood functions for a given $\theta$. Let $\mu$ be the counting measure on $\mathcal Y.$ For each $\theta$, define 
\begin{align}
	\mathfrak q_\theta=\{q_{y|x}:q_{y|x}=dQ(\cdot|x)/d\mu,Q\in \mathcal Q_\theta\}.
\end{align}
This set collects all (conditional) densities compatible with a given $\theta$. In the case of discrete games (Examples \ref{ex:ci_game}), this set contains all densities of equilibrium outcomes that are compatible with the game's description. Similarly, in the context of panel discrete choice (Example \ref{ex:paneldc}), this set collects all densities of individual choices consistent with arbitrary specifications of the initial condition.  
Observe that $\mathfrak q_\theta$ reduces to a singleton set $\{q_\theta\}$ if the model is complete, in which case $q_\theta=dQ_\theta/d\mu$ with $Q_\theta(A|x)=\int 1\{g(u|x;\theta)\in A\}dF_\theta(u)$.

While the multiplicity of likelihood functions may appear challenging, $\mathfrak q_\theta$ can be simplified, and this property also simplifies our tests.
By Artstein's inequality, we may rewrite $\mathfrak q_\theta$  as follows \citep[see e.g.][]{galichon2011set,molinari2020microeconometrics}:
\begin{align}
    \mathfrak q_\theta=\Big\{q_{y|x}:\sum_{y\in A}q_{y|x}(y|x)\ge \nu_\theta(A|x),~A\subseteq \mathcal Y\Big\},\label{eq:def_qset_theta}
\end{align}
where
\begin{align}
    \nu_\theta(\cdot|x)=F_\theta(G(u|x;\theta)\subseteq \cdot|x)\label{eq:def_belief}
\end{align} is the conditional \emph{containment functional} (or \emph{belief function}) associated with the random set $G(u|x;\theta)$. This function gives the 
sharp lower bound for the conditional probability $Q(A|x)$ across all $Q$'s belonging to $\mathcal Q_\theta$.\footnote{The upper bound for $Q(A|x)$ is given by the capacity functional $\nu^*(A|X)=F_\theta(G(u|x;\theta)\cap A\ne\emptyset|x)$. It is sufficient to use either of the lower or upper bounds in \eqref{eq:def_qset_theta} because the bounds are related to each other through the conjugate relationship $\nu(A|x)=1-\nu^*(A^c|x)$.} Theoretical properties of the containment functional and numerical approximation methods are well studied.\footnote{See \cite{molchanov2005theory} for a general treatment. For numerical approximations, see \cite{CilibertoTamer2009,galichon2011set}. We briefly review them in Appendix \ref{sec:notation}. }
For us, it is important that the linear inequalities in \eqref{eq:def_qset_theta} characterize $\mathfrak q_\theta$. Together with an extended Neyman-Pearson lemma reviewed below, this characterization makes the score computation feasible. In the following subsection, we briefly review the existing results we will rely on.

\subsection{Least Favorable Parametric Model}\label{ssec:prelim}
Let $p_0(y|x)$ denote the true conditional density of $Y$ given  $X$.
Consider distinguishing a parameter value $\theta_0$ from another value $\theta_1$ in a parametric  model $\{p_\theta,\theta\in\Theta\}$ of conditional densities. This amounts to testing a simple null hypothesis $p_0=p_{\theta_0}$ against a simple alternative hypothesis $p_0=p_{\theta_1}$.
By the Neyman-Pearson lemma, the most powerful test is the likelihood-ratio (LR) test. In incomplete models, corresponding null and alternative hypotheses would be $p_0\in\mathfrak q_{\theta_0}$ and $p_0\in\mathfrak q_{\theta_1}$ rendering both hypotheses composite.  \cite{kz} showed that it was possible to extend the Neyman-Pearson lemma to such settings, building on a general result by \cite{HS}. We briefly summarize their results below.

Let $\phi:\mathcal Y\times\mathcal X\to[0,1]$ be a test and $E_q[\phi(Y,X)]$ be its rejection probability, under conditional density $q$ and the marginal distribution of $X$.\footnote{The rejection probability can be written as $E_q[\phi(Y,X)]=E[E_q[\phi(Y,X)|X]]$. Only the conditional expectation depends on $q$.} For a given $\theta$, the power guarantee of $\phi$ is $\pi_{\theta}(\phi)\equiv\inf_{q\in\mathfrak q_{\theta}}E_q[\phi(Y,X)]$, which is the power value certain to be obtained regardless of the unknown selection mechanism. \cite{kz} seeked for a \emph{level-$\alpha$ minimax test} \citep[see e.g.][]{lehmann2005testing} such that
\begin{align}
    \sup_{q\in\mathfrak q_{\theta_0}}E_q[\phi(Y,X)]\le \alpha,\label{eq:level_const}
\end{align}
and
\begin{align}
    \pi_{\theta_1}(\phi)\ge \pi_{\theta_1}(\tilde\phi),~\forall \tilde\phi~\text{ satisfying \eqref{eq:level_const}}.
\end{align}
The minimax test is a procedure that maximizes the power guarantee among tests that meet the uniform size control requirement.

Results from \cite{HS} imply that, when $\mathfrak q_{\theta_0}\cap\mathfrak q_{\theta_1}=\emptyset$, the rejection region of a minimax test is of the form $\{(y,x):\Lambda(y,x)>t\}$ for a measurable function $\Lambda:\mathcal Y\times\mathcal X\to\mathbb R$. Furthermore,
there is a \emph{least favorable pair (LFP)} of densities $(q_{\theta_0},q_{\theta_1})\in \mathfrak q_{\theta_0}\times\mathfrak q_{\theta_1}$ such that for all $t\ge 0$,
\begin{align}
    E_{q_{\theta_0}}[1\{\Lambda(Y,X)>t\}]=\sup_{q\in \mathfrak q_{\theta_0}}E_{q}[1\{\Lambda(Y,X)>t\}],
\end{align}
and
\begin{align}
     E_{q_{\theta_1}}[1\{\Lambda(Y,X)>t\}]&=\inf_{q\in \mathfrak q_{\theta_1}}E_{q}[1\{\Lambda(Y,X)>t\}],
\end{align}
where $\Lambda(y,x)=q_{\theta_1}(y|x)/q_{\theta_0}(y|x)$.
That is, $q_{\theta_0}\in \mathfrak q_{\theta_0}$ is a density consistent with $\theta_0$ and least favorable for controlling the size of a test among all elements of $\mathfrak q_{\theta_0}$,  and $q_{\theta_1}\in\mathfrak q_{\theta_1}$ is a density consistent with $\theta_1$ and least favorable for maximizing a measure of power among all elements of $\mathfrak q_{\theta_1}$. \cite{kz} exploited these features to show that a level-$\alpha$ minimax test is an LR test based on the LFP:
\begin{align*}
    \phi(y,x)=\begin{cases}
    1&\Lambda(y,x)>C\\
    \gamma&\Lambda(y,x)=C\\
    0&\Lambda(y,x)<C,
    \end{cases}
\end{align*}
where $(C,\gamma)$ solves $E_{q_{\theta_0}}[\phi(Y,X)]=\alpha.$

One can also characterize the LFP $(q_{\theta_0},q_{\theta_1})$ as a solution to the following  convex program.
\begin{align}
   (q_{\theta_0},q_{\theta_1})= \argmin_{(q_0,q_1)}\sum_{y\in\mathcal Y}&\ln \Big(\frac{q_0(y|x)+q_1(y|x)}{q_0(y|x)}\Big)(q_0(y|x)+q_1(y|x))\label{eq:cvx_prog}\\
    s.t. &\sum_{y\in A}q_{0}(y|x)\ge \nu_{\theta_0}(A|x),~A\subseteq \mathcal Y\label{eq:cvx_prog1}\\
    &\sum_{y\in A}q_{1}(y|x)\ge \nu_{\theta_1}(A|x),~A\subseteq \mathcal Y.\label{eq:cvx_prog2}
\end{align}
The constraints in \eqref{eq:cvx_prog1} and \eqref{eq:cvx_prog2} are  the sharp identifying restrictions.\footnote{A common way to use them for identification analysis is to define the \emph{sharp identified set} as $\Theta_I=\{\theta:P(A|x)\ge \nu_\theta(A|x),a.s.\}.$ That is, given the conditional probability $P(\cdot|x)$ identified from data, one collects all values of $\theta$ satisfying the sharp identifying restrictions. For hypothesis testing, we instead fix $\theta$ and ask what would be a distribution among all distributions satisfying the sharp identifying restrictions, which is least favorable for controlling the size or maximizing the power.} In view of \eqref{eq:def_qset_theta}, they are equivalent to imposing the restrictions that $q_0$ belongs to $\mathfrak q_{\theta_0}$ and $q_1$ belongs to $\mathfrak q_{\theta_1}$ respectively. These restrictions are useful for computing the LFP because they are linear in $(q_0,q_1)$. Also, the objective function is strictly convex in $(q_0,q_1)$.
One can solve the convex program above numerically in general. For the examples we discussed earlier, it is also possible to compute the LFP analytically (see Appendix \ref{sec:example_detail}). 

To illustrate, let us consider Example \ref{ex:ci_game}. Suppose  that the latent payoff shifters $(U^{(1)},U^{(2)})$ follow a bivariate standard normal distribution.
We may then compute $\nu_\theta(A|x)$ for each event. Let us take $A=\{(1,0)\}$ as an example.
 Using \eqref{eq:strategic_sub} and \eqref{eq:def_belief}, we obtain
\begin{multline}
    \nu_\theta(\{(1,0)\}|x)=F_\theta(G(u|x;\theta)\subseteq \{(1,0)\}|x )\\
    =(1-\Phi_{2})\Phi_{1}+\Phi(x^{(1)}{}'\delta^{(1)}+\beta^{(1)})[\Phi_{2}-\Phi(x^{(2)}{}'\delta^{(2)}+\beta^{(2)})].
\end{multline}
The last expression corresponds to the probability assigned to the green region in Figure \ref{fig:example1} (right panel) and is the sharp lower bound for the probability of $A=\{(1,0)\}$.

Now consider two parameter values $\theta_0=(0'_2,\delta')'$ and $\theta_1=(\beta',\delta')'$, where $\beta=(\beta^{(1)},\beta^{(2)})'$ with $\beta^{(j)}<0$ for $j=1,2$.
As discussed in more detail below, the model is complete when $\beta=0.$ One can show that  \eqref{eq:cvx_prog1} reduces to the following equality restrictions:
\begin{align}
    q_{0}((0,0)|x)&=(1-\Phi(x^{(1)}{}'\delta^{(1)}))(1-\Phi(x^{(2)}{}'\delta^{(2)}))\label{eq:null00}\\
    q_{0}((0,1)|x)&=(1-\Phi(x^{(1)}{}'\delta^{(1)}))\Phi(x^{(2)}{}'\delta^{(2)})\label{eq:null01}\\
    q_{0}((1,0)|x)&=\Phi(x^{(1)}{}'\delta^{(1)})(1-\Phi(x^{(2)}{}'\delta^{(2)}))\label{eq:null10}\\
    q_{0}((1,1)|x)&=\Phi(x^{(1)}{}'\delta^{(1)})\Phi(x^{(2)}{}'\delta^{(2)}).\label{eq:null11}
\end{align}
They uniquely determine the least-favorable null density $q_{\theta_0}$ as follows
\begin{multline}
q_{\theta_0}(y|x)=[(1-\Phi_{1})(1-\Phi_{2})]^{1\{y=(0,0)\}}[(1-\Phi_{1})\Phi_{2}]^{1\{y=(0,1)\}}\\\times  [\Phi_{1}(1-\Phi_{2})]^{1\{y=(1,0)\}}[\Phi_{1}\Phi_{2}]^{1\{y=(1,1)\}},\label{eq:qtheta0}
\end{multline}
where,  to ease notation,  we use $\Phi_{1}$ and $\Phi_{2}$ to denote $\Phi(x^{(1)}{}'\delta^{(1)})$ and $\Phi(x^{(2)}{}'\delta^{(2)})$.

When $\beta^{(j)}<0,j=1,2$, there are multiple densities satisfying \eqref{eq:cvx_prog2}.
The  least favorable alternative density $q_{\theta_1}$ can be found by minimizing \eqref{eq:cvx_prog} with respect to $q_1$ subject to \eqref{eq:cvx_prog2}. The solution can be expressed analytically. For example, when player 1's strategic interaction effect on player 2 is relatively high, it is given by the following form:\footnote{Appendix \ref{sec:app_discretegame} gives a full characterization of the LFP for Example \ref{ex:ci_game}.  }
\begin{multline}
q_{\theta_1}(y|x)=[(1-\Phi_{1})(1-\Phi_{2})]^{1\{y=(0,0)\}}[(1-\Phi(x^{(1)}{}'\delta^{(1)}+\beta^{(1)}))\Phi_{2}]^{1\{y=(0,1)\}}\\
\times[(1-\Phi_{2})\Phi_{1}+\Phi(x^{(1)}{}'\delta^{(1)}+\beta^{(1)})(\Phi_{2}-\Phi(x^{(2)}{}'\delta^{(2)}+\beta^{(2)}))]^{1\{y=(1,0)\}}\\
\times[\Phi(x^{(1)}{}'\delta^{(1)}+\beta^{(1)})\Phi(x^{(2)}{}'\delta^{(2)}+\beta^{(2)})]^{1\{y=(1,1)\}}.\label{eq:qtheta1}
\end{multline}

Comparing \eqref{eq:qtheta0} and \eqref{eq:qtheta1}, one can see that $q_{\theta_1}$ tends to $q_{\theta_0}$ as $\beta$ approaches its null value (i.e. 0). Hence, one may view $\theta\mapsto q_{\theta}$ as a ``parametric'' model.
For each $\theta_1$, the density $q_{\theta_1}$ corresponds to the data generating process that is least favorable in detecting $\beta$'s deviation from its null value among all densities compatible with $\theta_1$. 
By varying $\theta_1$, we may trace out a family of such densities and form a parametric model. 
We define a \emph{least favorable (LF) parametric model} as follows.
\begin{definition}[Least favorable parametric model]
Let $\tilde\Theta\subseteq \Theta$ be an open set containing $\Theta_0$.
A family of densities $\{q_\theta:\theta\in\tilde\Theta\}$ is the \emph{least favorable (LF) parametric model} indexed by $\theta\in\tilde\Theta$ if (a) $q_\theta$ is the unique element of $\mathfrak q_\theta$ for any $\theta\in\Theta_0$; and (b) $q_\theta$ is the density of the least-favorable alternative distribution $Q_1$ if $\theta\in \tilde\Theta\setminus \Theta_0$, i.e., $Q_1$ constitutes a LFP $(Q_0,Q_1)\in \mathcal P_{\theta_0}\times\mathcal P_{\theta}$ with $\theta_0=(\beta_0',\delta')'$ and $\theta=(\beta',\delta')'$.
\end{definition}
Part (a) of the definition above is natural because the model is complete under the null hypothesis. Part (b) deserves discussion. We associate each alternative parameter value $\theta\not\in\Theta_0$ with a density that corresponds to the least favorable density for testing between $\theta_0=(\beta_0,\delta)$ and $\theta=(\beta,\delta)$. In other words, for each $\delta$, we focus on testing $\beta_0$ against $\beta$ using the least-favorable distribution for maximizing power. We construct $q_\theta$ this way because we aim at detecting local deviations in terms of $\beta$.
As we see below, this approach allows us to capture the sensitivity of $q_\theta$ with respect to a change in the parameter of interest through a score function.\footnote{An alternative choice of $\theta_0\in\Theta_0$ is also possible, which we do not seek here because it does not seem to lead to a tractable score test.} We provide a sufficient condition for the existence of an LF parametric model in the next section. 

Let us also note the following points. First, having a complete null model is not sufficient for obtaining a score function. It also requires us to define parametric densities at local alternatives. Our approach is to select the density associated with the least favorable selection mechanism for power maximization. Second, we do not need to know the precise form of the selection mechanism that induces $q_{\theta}$ (for $\theta\in \Theta_1$). Solving the convex program, we  ``profile out'' the selection mechanism and directly obtain the induced density $q_{\theta}$. This is why  $q_{\theta}$ is a function of $\theta$ only and does not involve any selection mechanism.

Coming back to equation \eqref{eq:qtheta1}, the formula suggests we may pretend as if data were generated by a parametric discrete choice model with the given density. Thanks to this feature, most of our analysis below will resemble that of standard discrete choice models.

\subsection{Model Completeness under the Null Hypothesis}\label{sec:completeness}
In this section, we start with an assumption to ensure that the LF parametric model is well-defined over a parameter set that contains the null parameter space $\Theta_0$ and its neighborhood. For this, let $\theta_{h}=(\beta_0'+h',\delta')'$, and let $\mathbb C_\epsilon$ denote an open cube centered at the origin with edges of length $2\epsilon$.
\begin{assumption}\label{as:hypotheses}
(i) Under any null parameter value $\theta_0=(\beta_0',\delta')'$ with $\delta\in\Theta_\delta$, the model makes a complete prediction so that $\mathfrak q_{\theta_0}=\{q_{\theta_0}\}$ is a singleton set; (ii) There exists $\epsilon>0$ such that the two sets $\mathfrak q_{\theta_0}$ and $\mathfrak q_{\theta_{h}}$ are disjoint for any $h\in (B_1-\beta_0)\cap \mathbb C_\epsilon.$
\end{assumption}
Assumption \ref{as:hypotheses} (i) holds whenever the model makes a complete prediction under the null hypothesis and is satisfied in the examples discussed in Section \ref{sec:examples}.
The model can be incomplete under the alternative hypothesis. Assumption \ref{as:hypotheses} (ii) requires  $\mathfrak q_{\theta_{h}}$ does not share any element with $\mathfrak q_{\theta_0}$. Under this condition, it is possible to detect local deviations from the null hypothesis regardless of the unknown selection mechanism. Such alternatives are \emph{robustly testable} in the sense that there exists a test that has nontrivial power against any distribution in $\mathfrak q_{\theta_{h}}$ \citep{kz}. For this condition, it suffices to have an event $A\subset\mathcal Y$ such that $q_{\theta_0}(A|x)<\nu_{\theta_{h}}(A|x)$ (or $q_{\theta_0}(A|x)>\nu_{\theta_{h}}^*(A|x)$) for all $\tau>0$ for some $x\in\mathcal X$.
All of the examples in Section \ref{sec:examples} satisfy Assumption \ref{as:hypotheses} (ii), and we demonstrate how to show this condition in Appendix \ref{sec:example_detail}. We construct score tests that have power against robustly testable local alternatives.\footnote{\cite{kz} extend the notion of local alternatives and analyze a more general setting that does not require Assumption \ref{as:hypotheses} (ii).  We conjecture that we may extend our framework similarly. Since all of our examples satisfy Assumption  \ref{as:hypotheses} (ii), we leave this extension elsewhere.}

Let us revisit the examples.
\setcounter{example}{0}
\begin{example}[Discrete Games of Strategic Substitution]\rm
Consider testing the presence of strategic substitution effects by testing $H_0:\beta^{(1)}=\beta^{(2)}=0$ against $H_0:\beta^{(1)}<0,\beta^{(2)}<0$. Under the null hypothesis, there is no strategic interaction between the players, which leads to the following complete prediction:
\begin{align}
    G(u|x;\theta_0)=
\begin{cases}
\{(0, 0)\} & u^{(1)}<-x^{(1)}{}'\delta^{(1)}, u^{(2)}<-x^{(2)}{}'\delta^{(2)},\\
\{(1, 1)\} & u^{(1)}>-x^{(1)}{}'\delta^{(1)}, u^{(2)}>-x^{(2)}{}'\delta^{(2)},\\
\{(1, 0)\} & u^{(1)}>-x^{(1)}{}'\delta^{(1)}, u^{(2)}\le -x^{(2)}{}'\delta^{(2)},\\
\{(0, 1)\} & u^{(1)}\le -x^{(1)}{}'\delta^{(1)}, u^{(2)}>-x^{(2)}{}'\delta^{(2)}.
\end{cases}\label{eq:ex1_complete}
\end{align}
Hence, for any value of the observed and unobserved variables,  $G(u|x;\theta_0)$ contains a unique equilibrium outcome (left panel of Figure \ref{fig:example1}). Combining the complete prediction with a parametric assumption on $U$ ensures Assumption \ref{as:hypotheses} (i)

We use this example to discuss Assumption \ref{as:hypotheses} (ii).
Under any $\theta_1$ with $\beta^{(j)}<0,j=1,2$, the probability allocated to $A=\{(1,1)\}$ is lower than that under the null as (right panel in Figure \ref{fig:example1}), provided $U$ is continuously distributed over $\mathbb R^2$. Hence, $\nu_{\theta_0}^*(\{(1,1)\}|x)<\nu_{\theta_1}(\{(1,1)\}|x)$, which ensures Assumption 1 (ii).
\end{example}

We also note that Examples 2-3 reduce to complete models under the null hypothesis.\footnote{To save space, we show Assumption \ref{as:hypotheses} (ii) for these examples in Appendix \ref{sec:example_detail}.}

\begin{example}[Triangular Models with a Set-valued Control Function]\rm
Consider testing the endogeneity of the treatment by testing the hypothesis that the coefficient $\beta$ on the control function $v$ is 0. When the null hypothesis is true,  the model's prediction reduces to 
\begin{align}
    y_{i}=1\{\alpha d_i+w_i'\eta+u_i\ge 0 \},~i=1,\dots,n.
\end{align}
 This is a  standard binary choice model with exogenous covariates.
For a given $(x_i,u_i)$, $y_i$ is uniquely determined. There is no need to control for $V_i$ because $U_i$ is independent of $(D_i,W_i)$. 
\end{example}

\begin{example}[Panel Dynamic Discrete Choice Models]\rm
Consider testing the presence of state dependence by testing whether the coefficient $\beta$ on the lagged dependent variable is 0. When $\beta=0$ in \eqref{eq:panel_ddc1}, the model reduces to a static panel binary choice model:
\begin{align}
    Y_{it}=1\{X_{it}'\eta+\alpha_i+\epsilon_{it}\ge 0\},~i=1,\dots, n, ~t=1,\dots,T, \label{eq:ex3_unique}
\end{align}
which makes
\eqref{eq:panel2}-\eqref{eq:panel3} and \eqref{eq:panel5}-\eqref{eq:panel6} equivalent. Under $H_0$, $G(u_i|x_i;\theta_0)$ contains the unique outcome value satisfying \eqref{eq:ex3_unique}.
\end{example}
We conclude this subsection with the following proposition.

\begin{proposition}\label{prop:lfmodel}
Suppose Assumption  \ref{as:hypotheses} holds. Then, a least-favorable parametric model $\{q_\theta:\theta\in\tilde\Theta\}$ exists for $\tilde\Theta=\{\theta=(\beta_0+h,\delta):h\in (B_1-\beta_0)\cap \mathbb C_\epsilon,\delta\in\Theta_\delta\}$.
\end{proposition}

\subsection{Score Tests}
Score-based tests such as Rao's score (or Lagrange multiplier) test and Neyman's $C(\alpha)$ test are widely used. They require the estimation of the restricted model only, which is particularly attractive in our setting. The restricted model is complete and typically admits point estimation of nuisance parameters under reasonably weak conditions. We take advantage of this property to carry out a score-based test.
Below, we briefly review the core ideas behind the classic score tests and discuss extensions to handle potential model incompleteness under the alternative.
For expositional purposes, we assume  $q_\theta$ is differentiable with respect to $\theta$ for now and will weaken this assumption later. 

Consider testing the null parameter value $\theta_0=(\beta_0',\delta')'$ against a local alternative hypothesis $\theta_h=(\beta_0'+h',\delta')'$, where  $h\in\mathbb R^{d_\beta}$. As discussed earlier, the optimal test in terms of guaranteed power is the likelihood-ratio test based on the LFP. The test is also robust in that, under Assumption \ref{as:hypotheses} (ii), the log-likelihood ratio can detect any deviation from the null hypothesis with non-trivial power regardless of the selection mechanism.

One can locally approximate the log-likelihood ratio
 by $\sum_{i=1}^n h's_{\beta}(Y_i|X_i;\beta_0,\delta)$, where $s_{\beta}(y|x;\beta,\delta)=\frac{\partial}{\partial \beta }\ln q_\theta(y|x)|_{\theta=(\beta,\delta)}$ is the score function. Let $\Sigma_{\beta_0}=\text{Var}(\sum_{i=1}^ns_{\beta}(Y_i|X_i;\beta_0,\delta))$. For i.i.d. data, $\Sigma_{\beta_0}=n I_{\beta_0}$ where $I_{\beta_0}=E[s_{\beta}(Y_i|X_i;\beta_0,\delta)s_{\beta}(Y_i|X_i;\beta_0,\delta)']$.
For a fixed $h$,  the normalized quantity  
\begin{align}
	\frac{(\sum_{i=1}^n h's_{\beta}(Y_i|X_i;\beta_0,\delta))^2}{h'\Sigma_\beta h}\label{eq:normalized_score}
\end{align}
 serves as a robust measure of discrimination between $\beta_0$ and $\beta_0+h$. 
It locally approximates the log-likelihood ratio $\ln (q_{\theta_h}/q_{\theta_0})$.
The direction $h$ that maximizes \eqref{eq:normalized_score} is $h^*=I_{\beta_0}^{-1}\frac{1}{\sqrt n}\sum_{i=1}^n s_{\beta}(Y_i|X_i;\beta_0,\delta)$, which motivates Rao's score statistic:\footnote{See \cite{Bera:2001va} for a more detailed argument for complete models. The same argument can be applied to incomplete models by replacing the standard likelihood function with the LF density $q_\theta$.}
\begin{align}
	T_n=\sup_{h\in\mathbb R^{d_\beta}} \frac{\sum_{i=1}^nh' s_{\beta}(Y_i|X_i;\beta_0,\delta))^2}{nh'I_{\beta_0}h}=\frac{1}{\sqrt n}\sum_{i=1}^n s_{\beta}(Y_i|X_i;\beta_0,\delta)'I_{\beta_0}^{-1}\frac{1}{\sqrt n}\sum_{i=1}^n s_{\beta}(Y_i|X_i;\beta_0,\delta).\label{eq:def_rao1}
\end{align}
 Suppose that the nuisance parameter $\delta$ can be estimated by a point estimator $\hat\delta_n$. 
Evaluating the sample mean of the score at $\delta=\hat\delta_n$ and imposing the null hypothesis yields
\begin{align}
g_n(\beta_0)=\frac{1}{\sqrt n}\sum_{i=1}^n s_\beta(Y_i|X_i;\beta_0,\hat\delta_n).
\end{align}
A feasible version of \eqref{eq:def_rao1} is
\begin{align}
 \hat T_n=   g_n(\beta_0)'\hat V_n^{-1}g_n(\beta_0),\label{eq:def_Tn}
\end{align}
where $\hat V_n$ is an  estimator of the asymptotic variance $V_0\equiv I_{\beta_0}$. For example, one can use the sample analog $\hat V_n= n^{-1}\sum_{i=1}^ns_\beta(Y_i|X_i;\beta_0,\hat\delta_n)s_\beta(Y_i|X_i;\beta_0,\hat\delta_n)'$ or its regularized version.\footnote{For example, the following estimator proposed by \cite{andrews_barwick12} ensures that it is always nonsingular and is equivariant to scale changes
\begin{align}
\hat V_n=\hat\Sigma_n+\max\{\epsilon-\det(\hat\Omega_n),0\}\hat D_n,~\epsilon=0.012,
\end{align}
where $\hat\Sigma_n=n^{-1}\sum_{i=1}^ns_\beta(Y_i|X_i;\beta_0,\hat\delta_n)s_\beta(Y_i|X_i;\beta_0,\hat\delta_n)'$, $\hat D_n=\text{diag}(\hat\Sigma_n)$, and $\hat\Omega_n=\hat D_n^{-1/2}\hat\Sigma_n\hat D_n^{-1/2}$.
}
Under regularity conditions, $\hat T_n$ converges in distribution to a $\chi^2$-distribution with $d_\beta$ degrees of freedom under the null hypothesis.

The analysis so far presumed that $q_\theta$ was differentiable, and $h\in\mathbb R^{d_\beta}$ was unrestricted. These assumptions may be restrictive in our context. For example, in discrete games of complete information,  the least favorable parametric model $h\mapsto q_{\theta_h}$ and its score can take different functional forms depending on whether the alternative hypothesis admits strategic substitution (i.e. $h<0$ as in Example \ref{ex:ci_game}) or strategic complementarity (i.e. $h>0$).  It is then natural to analyze these two cases separately.
Below, we weaken differentiability requirements to accommodate these features and allow the alternative hypothesis to be restricted (e.g., one-sided).

Recall that a set $\Gamma\subseteq\mathbb R^d$ is said to be locally equal to set $\Upsilon\subseteq\mathbb R^d$ if $\Gamma\cap \mathbb C_\epsilon=\Upsilon\cap \mathbb C_\epsilon$ for some $\epsilon>0$ \citep{Andrews:1999aa}.
\begin{assumption}[$L^2$-directional differentiability]\label{as:direc-diff}
(i)  $B_1-\beta_0$ is locally equal to a  convex cone $\mathcal V_1$; (ii)
For any	$\zeta\in \mathcal V_1\times \mathbb R^{d_\delta}$, there exists a square integrable function $s_{\theta}=(s_\beta',s_\delta')':\mathcal Y\times\mathcal X\to\mathbb R^d$ such that
\begin{align}
\Big\|q_{\theta_0+\tau \zeta}^{1/2}-q_{\theta_0}^{1/2}(1+\frac{1}{2}\tau \zeta's_{\theta}(\cdot|\cdot;\beta_0,\delta))\Big\|_{L^2_\mu}=o(\tau),\label{eq:direc_diff}
\end{align}
as $\tau\downarrow0$.
\end{assumption}
Assumption \ref{as:direc-diff} (i) requires the set of deviations (from $\beta_0$) can be locally approximated by a convex cone.
In Example \ref{ex:ci_game}, consider testing $H_0:\beta=(0,0)'$ against  $H_1:\beta^{(1)}<0,\beta^{(2)}<0$. Then,   $B_1-\beta_0$ is locally equal to 
\begin{align}
\mathcal V_1=\{h=(h^{(1)},h^{(2)}):h^{(1)}<0,h^{(2)}<0\}.
\end{align}
Assumption \ref{as:direc-diff} (ii) uses the notion of differentiability in quadratic mean \citep[see, e.g.][]{Van-der-Vaart:2000aa}, but it only requires that a unique score, in the sense of the $L^2$-derivative of the square-root density, exists for the set $\mathcal V_1$ of local deviations from the null hypothesis. This weaker assumption is appropriate for incomplete models, and $s_\theta$ can be derived from the least favorable parametric model similar to the standard parametric models (see Appendix \ref{sec:example_detail}).

To accommodate the one-sided nature of the alternative hypothesis, we define a test statistic by
\begin{align}
\hat S_n=	g_n(\beta_0)'\hat V_n^{-1} g_n(\beta_0)-\inf_{h\in \mathcal V_1}( g_n(\beta_0)-h)'\hat V_n^{-1}( g_n(\beta_0)-h).\label{eq:def_Sn}
\end{align}
This test statistic is a modification of \eqref{eq:def_Tn} and follows the construction in \cite{Silvapulle:1995tm}. It requires the same functions of data as $T_n$, but it is designed to direct power against the local alternatives in $\mathcal V_1$. If the alternative hypothesis is locally unrestricted, i.e., $\mathcal V_1=\mathbb R^{d_\beta}\setminus 0$, the test statistic reduces to $T_n$.

The asymptotic distribution of $\hat S_n$ is no longer a $\chi^2$-distribution. However, its critical value is easy to compute using simulations. Let
\begin{align}
	c_\alpha = \inf\{x\in\mathbb R:Pr(S\le x)\ge 1-\alpha\},\label{eq:def_calpha}
\end{align}
where
\begin{align}
	S\equiv Z'V_0^{-1}Z-\inf_{h\in \mathcal V_1}(Z-h)'V_0^{-1}(Z-h),~Z\sim N(0,V_0),\label{eq:defS}
\end{align}
which can be simulated by drawing $Z$ repeatedly from a zero mean multivariate normal distribution with estimated variance $\hat V_n.$

\subsection{Restricted Maximum Likelihood Estimator}\label{sec:est_delta}
Let $q_{\beta_0,\delta}$ be the conditional density of $Y_i$ given $X_i$. By Assumption \ref{as:hypotheses} (i), this density is unique.
A natural estimator of $\delta$ is the restricted maximum likelihood estimator (RMLE) $\hat\delta_n$, which  maximizes the log-likelihood function
\begin{align}
	\mathbb M_n(\delta)\equiv  \frac{1}{n}\sum_{i=1}^n \ln q_{\beta_0,\delta}(Y_i|X_i).\label{eq:def_Mn}
\end{align}
The complete model (under $H_0$) is often a standard discrete choice problem. Hence, one can use package software (e.g., R, Stata) to compute the RMLE. Let us revisit the examples.

\setcounter{example}{0}
\begin{example}[Discrete Games of Strategic Substitution]\rm
Under $H_0:\beta^{(1)}=\beta^{(2)}=0$, the model has a unique likelihood function as discussed in Section \ref{ssec:prelim}. The RMLE $\hat\delta_n$ maximizes
\begin{align*}
  \mathbb M_n(\delta)&=\sum_{i=1}^n \Big( 1\{Y_i=(0,0)\} \ln[(1-\Phi_{1,i})(1-\Phi_{2,i})]
  +1\{Y_i=(0,1)\}\ln [(1-\Phi_{1,i})\Phi_{2,i}]\\
  &\qquad+1\{Y_i=(1,0)\}\ln[\Phi_{1,i}(1-\Phi_{2,i})]+1\{Y_i=(1,1)\}\ln[\Phi_{1,i}\Phi_{2,i}]\Big),
\end{align*}
where $\Phi_{j,i}=\Phi(X_i^{(j)}{}'\delta^{(j)}),j=1,2$.
\end{example}

\begin{example}[Triangular Models with a Set-valued Control Function]\rm
When $\beta=0$, there is no correlation between the errors in the outcome and selection equations.
The outcome equation reduces to a binary choice model with exogenous covariates. If we assume $U_i\sim N(0,1)$, we obtain a probit model with
\begin{align}
   P(Y_i=1|D_i=d_i,W_i=w_i)= q_{\beta_0,\delta}(1|d_i,w_i,z_i)=\Phi(\alpha d_i+w_i'\eta).\label{eq:ex4_condprob_null}
\end{align}
One can compute the RMLE of $\delta=(\alpha,\eta')'$ using package software. Similarly, the selection equation is another binary choice model. One can estimate the coefficients on the instruments $Z$ similarly.
\end{example}

\begin{example}[Panel Dynamic Discrete Choice Models]\rm
A random effects probit model assumes $\alpha_i$ is independent of $X_i$ and follows $N(0,\gamma^2)$, and $\epsilon_{i1},\dots,\epsilon_{iT}$ are independent standard normal random variables. This specification yields the following conditional density function:
\begin{align}
    q_{\beta_0,\delta}(y_i|x_i)=\int \prod_{t=1}^T\Phi\big[(2y_{it}-1)(x_{it}'\eta+\gamma a)\big]\phi(a)da.\label{eq:ex4likelihood}
\end{align}
One can construct a simulated likelihood function based on \eqref{eq:ex4likelihood} to obtain a restricted MLE of $\delta=(\eta',\gamma)'$ \citep{train_2009}.
\end{example}

\subsection{Asymptotic Properties}
This section collects results on the asymptotic properties of the score test.  The proofs of all theoretical results are in Appendix \ref{sec:appdx_lemmas_proofs}. Throughout, we assume that $U^n=(U_1,\dots,U_n)$ is an independent and identically distributed (i.i.d.) sample drawn from $F_\theta$, and $X^n=(X_1,\dots,X_n)$ is also an i.i.d. sample following $q_X^n$.  
The joint distribution of the outcome sequence $Y^n=(Y_1,\dots,Y_n)\in\mathcal Y^n$ conditional on $X^n=x^n$ is not uniquely determined due to the potential incompleteness of the model. For $\theta\in\Theta$, the distribution belongs to the following set:
\begin{multline}
	\mathcal Q^n_\theta=\Big\{Q:Q(A|x^n)=\int_{U^n} p(A|u^n,x^n)dF^n_\theta(u),~\forall A\subseteq \mathcal Y^n,\\
	~\text{for some }p\in\Delta_{Y^n|X^n,u^n}~\text{ such that }p(G^n(u^n|x^n;\theta)|u^n,x^n)=1, a.s.\Big\},\label{eq:defQn}
\end{multline}
where $F^n_\theta$ denotes the joint law of $U^n$, and $G^n(u^n|x^n;\theta)=\prod_{i=1}^n G(u_i|x_i;\theta)$ is the Cartesian product of the set-valued predictions. We let $\mathcal P_{\theta}^n$ collect joint laws  of $(Y^n,X^n)$; each element $P^n$ of $\mathcal P_{\theta}^n$ is such that the conditional law of $Y^n$ given $X^n$ belongs to $\mathcal Q_{\theta}^n$, and the law of $X^n$  is $q^n_X$.
Assuming $U^n$ and $X^n$ are i.i.d. does not imply $Y^n$ is i.i.d. The set $\mathcal Q^n_\theta$, in general, contains dependent and heterogeneous laws because the behavior of the selection mechanism across experiments is unrestricted \citep[see][]{eks}. This feature does not create an issue for the size properties of our test because $\mathcal Q_\theta^n$ reduces to a single i.i.d. law under the null hypothesis. 

For the asymptotic properties of the RMLE, we also allow $\beta$ to be in a local neighborhood of $\beta_0$. For such settings, we provide conditions under which $\hat\delta_n$ is  $\sqrt n$-consistent.
Let $h\in \mathcal V_1$ and $\delta_0\in\Theta_\delta$. We assume data are generated from $P^n \in \mathcal Q^n_{\beta_0+h/\sqrt n,\delta_0}$. The null hypothesis corresponds to the setting with $h=0$.

Fixing $\beta=\beta_0$, one can view $q_{\beta_0,\delta}$ as the conditional density of $Y$ in a regular parametric  model, in which $\delta$ is the only unknown parameter. 
For each $\delta\in\Theta_\delta$, let $\M(\delta)\equiv E[\ln \q_{\beta_0,\delta}(Y_i|X_i)]$, where expectation is taken with respect to the conditional density $q_{\beta_0,\delta_0}$ and the distribution of $X.$ Let $\Mn(\delta)$ be the sample counterpart of $\M$ defined in \eqref{eq:def_Mn}. 

\begin{assumption}\label{as:consistency}
(i-a) There is a continuous function $M:\Theta_\delta\to\mathbb R_+$ such that $$\sup_{(y,x)\in\mathcal Y\times\mathcal X}|\ln q_{\beta_0,\delta}(y|x)|\le M(\delta),~\text{ for all }\delta\in\Theta_\delta;$$

(i-b) The map $\delta\mapsto \ln\q_{\beta_0,\delta}(y|x)$  is Lipschitz continuous uniformly in   $(y,x)$. That is,
\begin{align}
    \sup_{(y,x)\in\mathcal Y\times\mathcal X}\big|\ln q_{\beta_0,\delta}(y|x)-\ln q_{\beta_0,\delta'}(y|x)\big|\lesssim \|\delta-\delta'\|~\forall \delta,\delta'\in\Theta_\delta.
\end{align}

(i-c) $\delta\ne\delta_0\Rightarrow \q_{\beta_0,\delta}(y|x)\ne \q_{\beta_0,\delta_0}(y|x)$ with positive probability;
 
(ii)  $\Theta_\delta$ is a nonempty compact subset of Euclidean space; 

(iii) The restricted MLE $\hat\delta_n$ satisfies $\Mn(\hat\delta_n)\ge \inf_{\delta\in\Theta_\delta}\Mn(\delta)+r_n$, for some sequence  $\{r_n\}$.  For any $\epsilon>0$ and $\theta$ in a neighborhood of $\theta_0$, $\sup_{P^n\in\mathcal P^n_{\theta}}P^n(|r_n|>\epsilon)\to 0$. 
\end{assumption}
Assumption \ref{as:consistency} imposes sufficient conditions for identification and uniform law of large numbers standard in the literature.
We also assume the density of $F_\theta$ depends on $\theta$ smoothly. For this, for any integrable function $f$ defined on a measure space $(A,\mathfrak F,\zeta)$,  let $\|f\|_{L^1_\zeta}$ be the $L^1$-norm of $f$.
\begin{assumption}\label{as:ftheta_Lip}
For each $\theta\in\Theta$, $F_\theta$ is absolutely continuous with respect to a $\sigma$-finite measure $\zeta$ on $U$. The Radon-Nikodym density $f_\theta=dF_\theta/d\zeta$ satisfies
\begin{align}
    \|f_\theta-f_{\theta'}\|_{L^1_\zeta}\le C\|\theta-\theta'\|,~\forall \theta,\theta'\in \Theta,
\end{align}
for some $C>0.$
\end{assumption}
Finally, the following condition ensures the population objective function is locally well behaved so that its value is informative about $\delta_0$.  

\begin{assumption}\label{as:majorant}
(i) $\M$ is twice continuously differentiable at $\delta_0$;

(ii) The Hessian matrix 
\begin{align}
   H(\delta_0)=\frac{\partial^2}{\partial\delta\partial\delta'}\M(\delta)\Big|_{\delta=\delta_0}
\end{align}
is negative definite.
\end{assumption}
Under these assumptions, the restricted MLE $\hat\delta_n$ is $\sqrt n$-consistent.
\begin{proposition}\label{prop:rootn-consistency}
Suppose Assumptions \ref{as:hypotheses}-\ref{as:majorant} hold. Then,
\begin{align}
    \sqrt n\|\hat\delta_n-\delta_0\|=O_{P^n}(1),
\end{align}
uniformly in $P^n\in \mathcal P^n_{\theta_0+h/\sqrt n}.$
\end{proposition}

Below, let $P^n_0\in \mathcal P^n_{\theta_0}$ be the joint law of $(Y^n,X^n)$ under the null hypothesis.
Let $s_{\theta,j}$ be the $j$-th component of $s_\theta$. 
Let
\begin{align}
    \xi_{j,k}(y,x;\delta)=s_{\theta,j}(y|x;\beta_0,\delta)s_{\theta,k}(y|x;\beta_0,\delta),\label{eq:def_xi}
\end{align}
and let 
$\Xi=\big\{f:\mathcal Y\times\mathcal X\to\mathbb R|f(y,x)=\xi_{j,k}(y,x;\delta),~1\le j,k\le d,\delta\in\Theta_{\delta}\big\}.$
Next, we add a condition for the asymptotic distribution of $\hat S_n$ and consistent estimation of the asymptotic variance $V_0$.
\begin{assumption}\label{as:gc}
(i) $\frac{\partial}{\partial \delta}E[s_\beta(Y|X;\beta_0,\delta)]$ exists on a neighborhood of $\delta_0$;

(ii) $\sup_{f\in\Xi}\big|\frac{1}{n}\sum_{i=1}^n f(Y_i,X_i)-E_{P_0}[f(Y_i,X_i)]\big|=o_{P_0^n}(1),$ and $\delta\mapsto E[\xi_{j,k}(Y,X;\delta)]$ is continuous for any $1\le j,k\le d$.

(iii) $V_0$ is nonsingular.
\end{assumption}
Here we assume the expected score can be linearized and the elements of $\Xi$ obey a uniform law of large numbers.
Suppose $\hat S_n$ as defined in \eqref{eq:def_Sn}.
The following theorem shows that the test controls its asymptotic size.
\begin{theorem}\label{thm:size}
Suppose Assumptions \ref{as:hypotheses}-\ref{as:gc} hold. Let $c_\alpha$ be defined as in \eqref{eq:def_calpha}.
Then, for any $\alpha\in (0,1)$,
\begin{align}
	\lim_{n\to\infty}P^n_0(\hat S_n>c_\alpha)=\alpha.
\end{align}
\end{theorem}

\subsection{Inference on Parameters}\label{sec:inf_param}
In some applications, the ultimate goal may be to make inference on the underlying parameter, for example, to construct confidence intervals for components of $\theta$.  While we defer a formal analysis to future work, we suggest a hybrid procedure that aims at controlling the potential distortion of the model selection step, borrowing insights from the moment selection literature \citep{andrews2010inference,romano2014practical}.

Consider constructing confidence intervals for a component or linear combination $\gamma_0=p'\delta_0$ of $\delta_0$.\footnote{Since the null hypothesis pins $\beta$'s value down, it is natural to consider inference on the parameters that are estimated under both null and alternative hypotheses.}
Due to Proposition \ref{prop:rootn-consistency}, $\hat\gamma_n=p'\hat\delta_n$ is a $\sqrt n$-consistent estimator of $\gamma_0$ as long as the true value of $\beta$ is in a neighborhood of $\beta_0$ whose radius is of order $n^{-1/2}$. It would be natural to use such an estimator to construct a confidence interval for $\delta_0$ if the complete model is selected. A well-known challenge for such post-model selection inference is that a naive asymptotic approximation that disregards the model selection step may not be valid uniformly over a large class of data generating processes \citep{leeb2005model,andrews2009hybrid}.
Given this, we consider the following hybrid method. 

\noindent
Step 1: Compute $\hat S_n$ and $c_{n}=(\kappa_n\wedge 1)c_\alpha$, where $\kappa_n$ is a sequence of shrinkage factors that tends to 0 slowly, e.g. $\kappa_n=(\ln n)^{-1/2}$;

\bigskip
\noindent
Step 2: 
\begin{itemize}
    \item Reject $H_0:\beta=\beta_0$ if $\hat S_n>c_n$. Construct a \emph{robust confidence interval} for $\gamma_0$ using methods such as \cite{BCS,KMS};
    \item Do not reject $H_0:\beta=\beta_0$ if $\hat S_n\le c_n$. Construct the \emph{Wald confidence interval}  $[\hat\gamma_n-z_{\alpha/2} SE(\hat\gamma_n),~\hat\gamma_n+z_{\alpha/2} SE(\hat\gamma_n)]$, where $SE(\cdot)$ is the (estimated) standard error of its argument, and $z_{\alpha}$ is the $1-\alpha$ quantile of the standard normal distribution. 
\end{itemize}

The heuristic behind this procedure is as follows. First, we compare $\hat S_n$ to a critical value $c_n$ that tends to 0 slowly. For DGPs whose $\beta$ is outside local neighborhoods of $\beta_0$, we cannot ensure the asymptotic validity of the Wald confidence interval. In such settings, the procedure above uses a robust confidence interval asymptotically, which controls the asymptotic coverage probability. Since the critical value tends to 0, we use the Wald confidence interval only if $\beta$ is in a local neighborhood of $\beta_0$. The shrinkage factor $\kappa_n$, therefore, introduces a conservative distortion, which is expected to make the resulting confidence interval's coverage probability above its nominal over a wide range of $\beta$ values.

\section{Empirical Illustrations}

We illustrate the score test through two empirical applications.  

\subsection{Testing Strategic Interaction Effects}
The first application revisits the analysis of the airline industry by \cite{klinetamer2016bayesian}. We test the presence of strategic interaction effects between two types of firms: low-cost carriers (LCC) and other airlines (OA). Below, we briefly summarize the setup and refer to \cite{klinetamer2016bayesian} for details.
 A market is defined as trips between airports regardless of intermediate stops. The two types of firms, LCC and OA, decide whether or not to serve each market. The binary variable $y^{(\ell)}_i$ takes value 1 if airline $\ell\in\{\text{LCC, OA}\}$ serves market $i$. Airline $\ell$'s payoff in market $i$ equals
$$y_{i}^{(\ell)}(\delta^{cons}_{\ell}+\delta^{size}_{\ell}X_{i, size}+\delta^{pres}_{\ell}X^{(\ell)}_{i, pres}+\beta_{\ell}y^{(-\ell)}_{i}+u^{(\ell)}_{i}),$$
where $\beta_{\ell}$ captures the impact of the competitor's entry decision, $y^{(-\ell)}_{i}$. The airline-specific intercepts and observable covariates determine each firm's payoff. The covariates include the \textit{market size} $X_{i, size}$ and the \textit{market presence} $X^{(\ell)}_{i,pres}$. The market size $X_{i, size}$ is defined as the population at the endpoints of each trip. The latter variable $X^{(\ell)}_{i,pres}$ measures the presence of firm $\ell$ in market $i$ (see \cite{klinetamer2016bayesian} p.356 for its  definition).  This airline-and-market-specific variable shows up only in firm $\ell$'s payoff. The data come from the second quarter of the 2010 Airline Origin and Destination Survey (DB1B) and contain 7882 markets.\footnote{The data are available on Brendan Kline's \href{www.brendankline.com}{website}.}

Our hypothesis of interest is whether the LCCs and OAs compete strategically, which can be formulated as a one-sided test. The null hypothesis is $H_0:\beta_{LCC} = \beta_{OA} = 0$, and the alternative hypothesis is $H_1:\beta_{\ell}<0,\ell \in \{\text{LCC,OA}\}$.
Finally, the vector of coefficients $\delta=(\delta^{cons}_{LCC},\delta^{size}_{LCC},\delta^{pres}_{LCC},\delta^{cons}_{OA},\delta^{size}_{OA},\delta^{pres}_{OA})$ is the nuisance parameter in this model.  We estimate $\delta$ by the restricted MLE under the null hypothesis.

The value of the test statistic is 24.668. The 5\% critical value is 5.050. Hence, we reject the null hypothesis at the 5\% level. This result is consistent with the finding of \cite{klinetamer2016bayesian} whose credible sets for the strategic interaction effects $\beta_\ell,\ell\in\{\text{LCC,OA}\}$ do not contain the origin.
 Table \ref{tab:delta_est} reports the RMLE of the index coefficients. 
 The estimates suggest that the effect of market presence is larger for LCCs than other airlines, and the monopoly profits (captured by the constant terms) in a market with below-median size and below-median market presence are smaller for the LCCs. These observations are also consistent with \cite{klinetamer2016bayesian}'s findings, although we note that these estimates are obtained by imposing the restriction rejected by the score test.
 
\begin{table}[h]
    \begin{center}
    \begin{tabular}{cccccc}
    \hline\hline
           $\hat{\delta}^{pres}_{LCC}$ & $\hat{\delta}^{size}_{LCC}$ & $\hat{\delta}^{cons}_{LCC}$ & $\hat{\delta}^{pres}_{OA}$ & $\hat{\delta}^{size}_{OA}$ & $\hat{\delta}^{cons}_{OA}$ \\
    \hline
    1.643 & 0.795 & -2.084 & 0.388 & 0.440 & 0.338 \\
    \hline
    \end{tabular}
    \caption{Estimated values of $\delta$ under $H_0$}
    \label{tab:delta_est}
    \end{center}
  \footnotesize{  Note:  $X^{(\ell)}_{i,size}$ and $X^{(\ell)}_{i,pres}$ are treated as continuous variables on the unit interval when they are not discretized; $X^{(\ell)}_{i,size}$ and $X^{(\ell)}_{i,pres}$ are binary indicators of whether the original variables are above their median or not when they are discretized.}
\end{table}

\subsection{Testing the Endogeneity of Catholic School Attendance}
The second application concerns the causal effect of Catholic school attendance on academic achievements studied by \cite{AltonjiElderTaber2005}. 
Whether Catholic schools provide a better education than public ones is important for education policies, but the analysis is complicated by the concern that selection into Catholic schools is nonrandom. Using the framework in Example \ref{ex:incomplete_cf}, we examine the endogeneity of Catholic school attendance by testing if the coefficient on the control function is zero. 

The data source is a subset of the National Educational Longitudinal Survey of 1988 (NELS:88). We use a version of the data available from  \cite{WooldridgeText}. We refer to \cite{AltonjiElderTaber2005} for a detailed discussion of the data. The dependent variable $y_{i}$ is a binary variable indicating whether the student graduated from high school by the year 1994. The binary treatment $d_{i}$ indicates whether the student attended a Catholic high school. The vector of exogenous control variables $w_{i}$ includes each parent's years of education and log family income. The instrument variable is a dummy variable indicating whether a parent was reported to be Catholic. The sample size $n$ is 5970 after we remove missing observations on $y_{i}$. 

Table \ref{tab:delta_est 2nd app} reports the point estimates of nuisance parameters $\delta$ under the null hypothesis $H_0:\beta=0$. The value of the test statistic is 154.848. The 5\% critical value is 2.755.  We, therefore, reject the null hypothesis at the 5\% level. Our test provides strong evidence supporting the students' selection into Catholic schools based on their unobservable characteristics. This result is in line with the concern expressed in \cite{AltonjiElderTaber2005}.
\begin{table}[h]
    \begin{center}
    \begin{tabular}{cccccccc}
    \hline\hline
          $\widehat{\alpha}$ & $\widehat{\eta}_{\text{motheduc}}$ & $\widehat{\eta}_{\text{fatheduc}}$ & $\widehat{\eta}_{\text{lfaminc}}$ & $\widehat{\gamma}_{\text{parcath}}$ & $\widehat{\gamma}_{\text{motheduc}}$ & $\widehat{\gamma}_{\text{fatheduc}}$ & $\widehat{\gamma}_{\text{lfaminc}}$\\
    \hline
    0.630 & 0.029 & 0.062 & 0.028 & 1.220 & 0.003 & 0.082 & -0.319 \\
    \hline
    \end{tabular}
    \caption{Estimated values of $\delta$ under $H_0$}
    \label{tab:delta_est 2nd app}
    \end{center}
  \footnotesize{  Note: Constants are not included in either equations (2.5) or (2.6). The variables listed in the table denote the mother's years of education, the father's years of education, the log family income, and whether one of the parents is reported to be Catholic. The estimation was conducted using the \textit{glmfit} in MATLAB.}
\end{table}

\section{Monte Carlo Experiments}
\subsection{Size and Power of the Score Test}
We examine the size and power properties of the score test through simulations. The data generating process is based on Example \ref{ex:ci_game} and is motivated by the empirical illustration in the previous section. There are player-specific covariates $X_i=(X_i^{(1)},X_i^{(2)})'$, each of which is generated as an independent Rademacher random variable taking values on $\{-1,1\}$.
We then generate  $U_i = (U^{(1)}_i, U^{(2)}_i)$ from the bivariate standard normal distribution. For each $u_i$ and $x_i$, we determine the predicted set of outcomes $G(u_i|x_i;\theta)$ based on the payoff functions with $\delta_0=(\delta_0^{(1)},\delta_0^{(2)})=(2,1.5)'$. We then test
\begin{align}
    H_0:\beta^{(1)}=\beta^{(2)}=0,~~~v.s.~~~H_1:\beta^{(1)}<0,\beta^{(2)}<0.
\end{align}
As discussed earlier, the model is complete under $H_0$. We estimate $\delta_0$ using the restricted MLE. The sample size is set to 2500, 5000, or 7500. This choice is motivated by the sample size used in the empirical application.

The size of the score test is reported in Table \ref{tab:size}. The size of the test is controlled properly across all sample sizes, while it tends to be slightly conservative when $n$ is small. 
\begin{table}[htbp]
    \centering
    \begin{tabular}{l|ccc}
      \hline\hline
      Sample size   & 2500 & 5000 & 7500  \\
      \hline
      Size & 0.028 & 0.036 & 0.041  \\
      \hline\hline
    \end{tabular}
    \caption{Size of the score test}
    \label{tab:size}
\end{table}
Under alternative hypotheses,  multiple equilibria may be predicted. If this is the case,   we select an outcome according to one of the following selection mechanisms. The first design uses a selection mechanism, which selects $(1, 0)$ out of $G(u_i|x_i;\theta)=\{(1, 0), (0, 1)\}$ if an i.i.d. Bernoulli random variable $\nu_{i}$ takes 1.
In the second design, we generate data from the least favorable distribution, which draws an independent outcome sequence from the least favorable distribution $Q_{\theta_1}\in \mathcal Q_{\theta_1}$.

The power of the score test is calculated against local alternatives with $\beta^{(j)}_1=-h/\sqrt n,h>0$ for $j=1,2.$ For this exercise, we introduce a grid of values for $h$ and generate the data described above. We then compare the rejection frequency of our test to that of the moment-based testing procedure by \cite{BCS}.  Their test checks if a hypothesized value $(\beta^{(1)},\beta^{(2)})'=(0,0)'$ is compatible with a set of moment restrictions.
Their statistic and bootstrap critical value are calculated using a sample analog of the following moment inequality and equality restrictions
\begin{align*}
   P(Y=(1,0)|X=x)&\ge (1-\Phi_{2})\Phi_{1}+\Phi(x^{(1)}{}'\delta_{1}+\beta^{(1)})[\Phi_{2}-\Phi(x^{(2)}{}'\delta_{2}+\beta^{(2)})]\\ 
   P(Y=(1,0)|X=x)&\le (1-\Phi(x^{(2)}{}'\delta_{2}+\beta^{(2)}))\Phi_{1}\\
   P(Y=(0,0)|X=x)&= (1-\Phi_{1})(1-\Phi_{2})\\
   P(Y=(1,1)|X=x)&=\Phi(x^{(1)}{}'\delta_{1}+\beta^{(1)})\Phi(x^{(2)}{}'\delta_{2}+\beta^{(2)}),
\end{align*}
which are the sharp identifying restrictions that characterize $\mathfrak q_\theta$ in \eqref{eq:def_qset_theta}.\footnote{Since the example resembles the specification used in their Monte Carlo experiments, we added minimal changes to their replication code posted on the repository of Quantitative Economics to implement their procedure.} 

Figures \ref{fig:power_iid}-\ref{fig:power_LFP} show the rejection frequencies of the score and moment-based tests. The results are similar across the two designs. In each design, the score test outperforms the moment-based test in terms of power by a significant margin. This difference in performance may potentially be due to the proposed score test's exploitation of completeness under the null to estimate the nuisance parameter and direct power against $\mathcal V_1$.
The moment-based test is designed for general subvector inference and does not necessarily exploit the model completeness.\footnote{The procedure by \cite{BCS} tests if the null parameter value is consistent with the model restrictions and deals with nuisance parameters by a profiling method combined with regularization to ensure its uniform validity.} The simulation results suggest that taking advantage of the model structure may provide considerable benefits in terms of power.

\section{Concluding Remarks}
Economic models exhibit incompleteness for various reasons. They are, for example, consequences of strategic interaction, state dependence, or self-selection. This paper shows that one can test these important features using a score statistic even if the model is incomplete under the alternative hypothesis. The proposed test exploits the model completeness under the null hypothesis to simplify its computation.
An avenue for future research includes a theory for the uniform validity of inference for post-model selection procedures based on the score test.

\clearpage
\bibliographystyle{ecta}

\clearpage
\appendix
\renewcommand{\baselinestretch}{0.9}
\small

\section{Notation and Preliminaries}\label{sec:notation}
The following list includes notation and definitions that will be used throughout the Appendices:
\begin{table}[h]
    \centering
    \begin{tabular}{rl}
        $a\lesssim b$   & $a\le Mb$ for some constant $M$.  \\
        $\|\cdot\|_{L^2}$ & the $L^2$-norm for square-integrable functions.\\
        $\|\cdot\|_{\mathcal F}$ & the supremum norm over $\mathcal F$.\\
        $N(\epsilon,\mathcal F,\|\cdot\|)$ & covering number of size $\epsilon$ for $\mathcal F$ under norm $\|\cdot\|$. \\
        $N_{[]}(\epsilon,\mathcal F,\|\cdot\|)$ & bracketing number of size $\epsilon$ for $\mathcal F$ under norm $\|\cdot\|$.\\
        $X_n\stackrel{P^n}{\leadsto}X$ & $X_n$ weakly converges to $X$ under $\{P^n\}$
    \end{tabular}
    \caption{Notation and definitions}
    \label{tab:notation}
\end{table}

Let $\Omega$ be a compact metric space and let $\Sigma_\Omega$ denote its Borel $\sigma$-algebra. Let $\mathcal K(\Omega)$ be the set of compact subsets of $\Omega$ endowed with the Hausdorff metric $d_H$. Let $\mathcal C(\Omega)$ be the set of continuous functions on $\Omega$. Let $\Delta(\Omega)$ be the set of  Borel probability measures on $\Omega$ endowed with the weak topology.

\subsection{Capacity functionals}
We briefly summarize the definition and properties of capacities.
A set function $\nu^{*}$ is said to be a \emph{capacity} if $\nu^{*}$ satisfies the following conditions:
\begin{enumerate}[label=(\roman*)]\label{def:capa}
	\item{$\nu^{*}(\emptyset)=0, \nu^{*}(\Omega)=1$,\label{cd:capac1}}
	\item{$A\subset B \Rightarrow \nu^{*}(A) \leq \nu^{*}(B)$,~ for all $A,B\in \Sigma_\Omega$.\label{cd:capac2}}
	\item{$A_n \uparrow A \Rightarrow \nu^{*}(A_n) \uparrow \nu^{*}(A)$, for all $\{A_n,n\ge 1\}\subset \Sigma_\Omega$ and $A\in\Sigma_\Omega$.\label{cd:capac3}}
	\item{$F_n \downarrow F, F_n$ closed $\Rightarrow \nu^{*}(F_n) \downarrow \nu^{*}(F)$.\label{cd:capac4}}
\end{enumerate}
One may define integral operations with respect to capacities as follows. Let $f:\Omega\to\mathbb R$ be a measurable function. The \emph{Choquet integral} of $f$ with respect to $\nu$ is defined by
\begin{align}
\int fd\nu\equiv \int_{-\infty}^0(\nu(\{\omega:f(\omega)\ge t\})-\nu(\Omega))dt+\int_0^\infty \nu(\{\omega:f(\omega)\ge t\})dt,
\end{align}
where the integrals on the right-hand side are Riemann integrals.
A capacity $\nu$ is said to be \emph{monotone of order $k$} or, for short, \emph{k-monotone} if for any $A_i\subset S,i=1\cdots,k$,
\begin{align}
	\nu\big(\cup_{i=1}^k A_i\big) \ge \sum_{I\subseteq\{1,\cdots,k\}, I\ne \emptyset}(-1)^{|I|+1}\nu\big(\cap_{i\in I}A_i\big).\label{eq:kmonotone}
\end{align}
Conjugate $\nu^*(A)=1-\nu(A^c)$ is then called a \emph{$k$-alternating} capacity. A capacity that satisfies \eqref{eq:kmonotone} is called an \emph{infinitely monotone capacity} or a \emph{belief function}. Capacities are used in various areas of statistics \citep{dempster67,shafer1976mathematical,wasserman1990belief} and economics \citep{GILBOA1989141}.

The following result, known as Choquet's theorem,  states that a random closed set $K$ following a distribution $M$ induces a belief function, and it follows from Theorems 1-3 in \cite{philippe1999decision}.
\begin{lemma}\label{lem:choquet}
	Let $\Omega$ be a Polish space. Let $M$ be a probability measure on $\mathcal K(\Omega)$. Let $\mathcal P=\{P\in\Delta(\Omega):P= \int P_KdM(K),P_K\in \Delta(K)\}$. Then, $\nu(\cdot)=\inf_{P\in\mathcal P}P(\cdot)$ is a belief function and satisfies
	\begin{align}
		\nu(A)=M(\{K\subseteq A\}).
	\end{align}
\end{lemma}
We apply the lemma above in our setting with a random subset of $\mathcal Y\times\mathcal X$.  Namely, we take $K=G(u|X;\theta)\times\{X\}$, and $M$ is the law of $K$ induced by $U$'s conditional distribution $F_\theta$ and $X$'s marginal distribution $q_x$. We then denote the induced belief function by $\nu_\theta(\cdot)$ and its conjugate $\nu^*_\theta(\cdot)=1-\nu_\theta(\cdot)$ (see \eqref{eq:def_nutheta}-\eqref{eq:def_nuthetastar} below).

\section{Details on the Examples}\label{sec:example_detail}
This section provides details on each of the examples discussed in the text. We present the sharp identifying restrictions, the least favorable parametric models, and primitive conditions for Assumption \ref{as:hypotheses}.
\subsection{Discrete Games of Complete Information}\label{sec:app_discretegame}

\subsubsection{Sharp Identifying Restrictions and Assumption \ref{as:hypotheses}}
Recall that $\Phi_j=\Phi(x^{(j)}{}'\delta^{(j)}),j=1,2$. The upper and lower probabilities of all singleton events are tabulated in  Table \ref{table:lowerprob}. In this example, they constitute the sharp identifying restrictions \citep{galichon2011set}.

\begin{table}[htbp]
	\centering 
	\footnotesize
	\caption{Upper and Lower Probability Bounds for Example 1}
	\begin{tabular}{c c c }
		\hline\hline
		 $A$ & $\nu_{\theta}(A)=\min P(A)$ &  $\nu^{*}_{\theta}(A)=\max P(A)$ \\ [0.5ex]
		\hline 
		\{(0, 0)\} & $(1-\Phi_{1})(1-\Phi_{2})$ & $(1-\Phi_{1})(1-\Phi_{2})$ \\ [1.2ex]
		\{(1, 1)\} & $\Phi(x^{(1)}{}'\delta^{(1)}+\beta^{(1)})\Phi(x^{(2)}{}'\delta^{(2)}+\beta^{(2)})$ & $\Phi(x^{(1)}{}'\delta^{(1)}+\beta^{(1)})\Phi(x^{(2)}{}'\delta^{(2)}+\beta^{(2)})$ \\ [1.2ex]
		\{(1, 0)\} & $(1-\Phi_{2})\Phi_{1}+\Phi(x^{(1)}{}'\delta^{(1)}+\beta^{(1)})[\Phi_{2}-\Phi(x^{(2)}{}'\delta^{(2)}+\beta^{(2)})]$ & $(1-\Phi(x^{(2)}{}'\delta^{(2)}+\beta^{(2)}))\Phi_{1}$ \\ [1.2ex]
		\{(0, 1)\} & $(1-\Phi_{1})\Phi_{2}+\Phi(x^{(2)}{}'\delta^{(2)}+\beta^{(2)})[\Phi_{1}-\Phi(x^{(1)}{}'\delta^{(1)}+\beta^{(1)})]$ & $(1-\Phi(x^{(1)}{}'\delta^{(1)}+\beta^{(1)}))\Phi_{2}$ \\ [1.2ex]
		\hline\hline 
	\end{tabular}
	\label{table:lowerprob} 
\end{table}

As argued in Section \ref{sec:completeness}, the model's prediction reduces to \eqref{eq:ex1_complete} when $\beta^{(1)}=\beta^{(2)}=0$, which implies a unique density in \eqref{eq:qtheta0}. Therefore, Assumption \ref{as:hypotheses} (i) holds. 
For Assumption \ref{as:hypotheses} (ii), it suffices to show that $\mathcal Q_{\theta_0}$ and $\mathcal Q_{\theta_1}$ are disjoint.
For this, consider the event $\{(1,1)\}$. Table \ref{table:lowerprob} suggests
\begin{align}
    \nu_{\theta_0}(\{(1,1)\}|x)=\Phi(x^{(1)}{}'\delta^{(1)})\Phi(x^{(2)}{}'\delta^{(2)}),
\end{align}
whereas
\begin{align}
    \nu_{\theta_1}^*(\{(1,1)\}|x)=\Phi(x^{(1)}{}'\delta^{(1)}+\beta^{(1)})\Phi(x^{(2)}{}'\delta^{(2)}+\beta^{(2)}).
\end{align}
This means $ \nu_{\theta_1}^*(\{(1,1)\}|x)<\nu_{\theta_0}(\{(1,1)\}|x)$ whenever $\beta^{(j)}<0,j=1,2$. Hence, $\mathcal Q_{\theta_0}$ and $\mathcal Q_{\theta_1}$ are disjoint.

\subsubsection{Least Favorable Parametric Model}
The least favorable parametric model $\theta\mapsto q_\theta$ is given by
\begin{align} 
q_{\theta}((0, 0)|x) & = (1-\Phi_{1})(1-\Phi_{2})\label{eq:LF_density1}\\
q_{\theta}((1, 1)|x) & = \Phi(x^{(1)}{}'\delta^{(1)}+\beta^{(1)}) \Phi(x^{(2)}{}'\delta^{(2)}+\beta^{(2)})\label{eq:LF_density2}\\
q_{\theta}((1, 0)|x) & =\begin{cases} \Phi_{1}(1-\Phi_{2})+\frac{\Phi_{1}\Phi_{2}-\Phi(x^{(1)}{}'\delta^{(1)}+\beta^{(1)})\Phi(x^{(2)}{}'\delta^{(2)}+\beta^{(2)})}{\Phi_{1}+\Phi_{2}-2\Phi_{1}\Phi_{2}}&\theta\in\Theta_1(x)\\
\Phi_{1}(1-\Phi_{2})+\Phi(x^{(1)}{}'\delta_{1}+\beta^{(1)})[\Phi_{2}-\Phi(x^{(2)}{}'\delta_{2}+\beta^{(2)})]&\theta\in\Theta_2(x)\\
 \Phi_{1}(1-\Phi(x^{(2)}{}'\delta^{(2)}+\beta^{(2)}))&\theta\in\Theta_3(x)\label{eq:LF_density3}
\end{cases},
\end{align} 
and $q_\theta((0,1)|x)$ is determined by $1-q_\theta((0,0)|x)-q_\theta((1,0)|x)-q_\theta((1,1)|x)$.
The parameter subsets, $\Theta_j(x),j=1,2,3$, are given by
\begin{align}
    \Theta_1(x)=\big\{\theta:&\Phi_{1}(1-\Phi(x^{(2)}{}'\delta^{(2)}+\beta^{(2)}))\ge\frac{z_{1}z_{2}-\Phi_{2}(1-\Phi_{1})z_{1}}{\Phi_{2}+\Phi_{1}-2\Phi_{1}\Phi_{2}}\\
&\frac{z_{1}z_{2}-\Phi_{2}(1-\Phi_{1})z_{1}}{\Phi_{2}+\Phi_{1}-2\Phi_{1}\Phi_{2}}\ge\Phi_{1}(1-\Phi_{2})+\Phi(x^{(1)}{}'\delta^{(1)}+\beta^{(1)})[\Phi_{2}-\Phi(x^{(2)}{}'\delta^{(2)}+\beta^{(2)})]\big\}\notag\\
 \Theta_2(x)=\big\{\theta:& \frac{z_{1}z_{2}-\Phi_{2}(1-\Phi_{1})z_{1}}{\Phi_{2}+\Phi_{1}-2\Phi_{1}\Phi_{2}}<\Phi_{1}(1-\Phi_{2})+\Phi(x^{(1)}{}'\delta^{(1)}+\beta^{(1)})[\Phi_{2}-\Phi(x^{(2)}{}'\delta^{(2)}+\beta^{(2)})]\big\}\label{eq:def_Theta2}\\
 \Theta_3(x)=\big\{\theta:&\Phi_{1}(1-\Phi(x^{(2)}{}'\delta^{(2)}+\beta^{(2)}))<\frac{z_{1}z_{2}-\Phi_{2}(1-\Phi_{1})z_{1}}{\Phi_{2}+\Phi_{1}-2\Phi_{1}\Phi_{2}}\big\},\label{eq:def_Theta3}
\end{align}
where
\begin{align*}
z_{1}&=\Phi_{1}(1-\Phi_{2})\\
z_{2}&=\Phi_{2}(1-\Phi_{1})+\Phi_{1}+\Phi_{2}-\Phi_{1}\Phi_{2}-\Phi(x^{(1)}{}'\delta^{(1)}+\beta^{(1)})\Phi(x^{(2)}{}'\delta^{(2)}+\beta^{(2)}).
\end{align*}
Below, we outline how to obtain this density from the convex program in \eqref{eq:cvx_prog}-\eqref{eq:cvx_prog2}.

As discussed in the text, $q_{\theta_0}$ is determined by the four equality restrictions \eqref{eq:null00}-\eqref{eq:null11}. Therefore, it remains to solve the convex program in \eqref{eq:cvx_prog}-\eqref{eq:cvx_prog2} for $q_1$. For this, we can reduce the number of control variables. First, Table \ref{table:lowerprob} implies 
\begin{align}
    q_{\theta_1}(0, 0|x)&=(1-\Phi_{1})(1-\Phi_{2})\\
    q_{\theta_1}(1, 1|x)&=\Phi(x^{(1)}{}'\delta^{(1)}+\beta^{(1)})\Phi(x^{(2)}{}'\delta^{(2)}+\beta^{(2)}).
\end{align}
Hence, the remaining free components of $q_1$ are $q_{1}(1, 0|x)$ and $q_{1}(0, 1|x)$. Let $\omega=q_{1}(1, 0|x)$. We may then express the other component as
$$q_{1}(0, 1|x)=1-q_{1}(0, 0|x)-q_{1}(1, 1|x)-\omega=\Phi_{1}+\Phi_{2}-\Phi_{1}\Phi_{2}-\Phi(x^{(1)}{}'\delta^{(1)}+\beta^{(1)})\Phi(x^{(2)}{}'\delta^{(2)}+\beta^{(2)})-\omega.$$ Hence, to solve \eqref{eq:cvx_prog}-\eqref{eq:cvx_prog2}, it suffices to choose $\omega=q_{1}(1, 0|x)$ optimally in the following problem:
\begin{align}
  \min_{\omega\in [0,1]} & -\ln\Big(\frac{z_{1}}{z_{1}+\omega}\Big)(z_{1}+\omega)-\ln\Big(\frac{(1-\Phi_{1})\Phi_{2}}{z_{2}-\omega}\Big)(z_{2}-\omega)\\
s.t. &~\omega-(1-\Phi(x^{(2)}{}'\delta^{(2)}+\beta^{(2)}))\Phi_{1}\le 0\notag\\
&(1-\Phi_{2})\Phi_{1}+\Phi(x^{(1)}{}'\delta^{(1)}+\beta^{(1)})[\Phi_{2}-\Phi(x^{(2)}{}'\delta^{(2)}+\beta^{(2)})]-\omega\le 0.\notag
\end{align}
Let the Lagrangian be
\begin{equation*}
\begin{split}
\mathcal{L}(\omega, \lambda)&=-\ln\Big(\frac{z_{1}}{z_{1}+\omega}\Big)(z_{1}+\omega)-\ln\Big(\frac{(1-\Phi_{1})\Phi_{2}}{z_{2}-\omega}\Big)(z_{2}-\omega)-\lambda_{1}((1-\Phi(x^{(2)}{}'\delta^{(2)}+\beta^{(2)}))\Phi_{1}-\omega)\\
&-\lambda_{2}(\omega-(1-\Phi_{2})\Phi_{1}-\Phi(x^{(1)}{}'\delta^{(1)}+\beta^{(1)})[\Phi_{2}-\Phi(x^{(2)}{}'\delta^{(2)}+\beta^{(2)})]).
\end{split}
\end{equation*}
The Karush-Kuhn-Tucker (KKT) conditions are
\begin{align}
&-\ln \bigg(\frac{z_{1}}{z_1+\omega}\bigg)+\ln
\bigg(\frac{\Phi_{2}(1-\Phi_{1})}{z_2-\omega}\bigg)+\lambda_1-\lambda_2=0\label{eq:KKT1}\\
&\lambda_1 \Big(\Phi_{1}(1-\Phi(x^{(2)}{}'\delta^{(2)}+\beta^{(2)}))-\omega\Big)\geq0\label{eq:KKT2}\\
&\lambda_{2}\Big(\omega-(1-\Phi_{2})\Phi_{1}-\Phi(x^{(1)}{}'\delta^{(1)}+\beta^{(1)})[\Phi_{2}-\Phi(x^{(2)}{}'\delta^{(2)}+\beta^{(2)})]\Big)\geq 0\label{eq:KKT3}\\
&\lambda_1,\lambda_2\geq 0.\label{eq:KKT4}
\end{align}	
Below, we consider three subcases depending on the value of the Lagrange multipliers.

\noindent\textbf{Case 1} ($\lambda_{1}=\lambda_{2}=0$):  The FOC in \eqref{eq:KKT1} with $\lambda_1=\lambda_2=0$ identifies the solution $q_{\theta_1}(1,0|x)$ as follows:
\begin{align}\omega=q_{\theta_1}(1,0|x)&=\frac{z_{1}z_{2}-\Phi_{2}(1-\Phi_{1})z_{1}}{\Phi_{2}+\Phi_{1}-2\Phi_{1}\Phi_{2}}\notag\\
&=\frac{\Phi_{1}(1-\Phi_{2})[\Phi_{1}+\Phi_{2}-\Phi_{1}\Phi_{2}-\Phi(x^{(1)}{}'\delta^{(1)}+\beta^{(1)})\Phi(x^{(2)}{}'\delta^{(2)}+\beta^{(2)})]}{\Phi_{2}+\Phi_{1}-2\Phi_{1}\Phi_{2}}.\label{eq:lfp_case1_10}
\end{align}
This implies 
\begin{align}
q_{\theta_1}(0, 1|x) &=\Phi_{1}+\Phi_{2}-\Phi_{1}\Phi_{2}-\Phi(x^{(1)}{}'\delta^{(1)}+\beta^{(1)})\Phi(x^{(2)}{}'\delta^{(2)}+\beta^{(2)})-\omega\\
&=\frac{\Phi_{2}(1-\Phi_{1})[\Phi_{1}+\Phi_{2}-\Phi_{1}\Phi_{2}-\Phi(x^{(1)}{}'\delta^{(1)}+\beta^{(1)})\Phi(x^{(2)}{}'\delta^{(2)}+\beta^{(2)})]}{\Phi_{2}+\Phi_{1}-2\Phi_{1}\Phi_{2}}.\label{eq:lfp_case1_01}
\end{align}
Substituting the value of $\omega$ into its bounds, we obtain the following restrictions:
\begin{align} &\Phi_{1}(1-\Phi(x^{(2)}{}'\delta^{(2)}+\beta^{(2)}))-\frac{z_{1}z_{2}-\Phi_{2}(1-\Phi_{1})z_{1}}{\Phi_{2}+\Phi_{1}-2\Phi_{1}\Phi_{2}}\geq0\label{eq:case1_res1}\\
& \frac{z_{1}z_{2}-\Phi_{2}(1-\Phi_{1})z_{1}}{\Phi_{2}+\Phi_{1}-2\Phi_{1}\Phi_{2}}-\Phi_{1}(1-\Phi_{2})-\Phi(x^{(1)}{}'\delta^{(1)}+\beta^{(1)})[\Phi_{2}-\Phi(x^{(2)}{}'\delta^{(2)}+\beta^{(2)})]\geq0.\label{eq:case1_res2}
\end{align}
We let $\Theta_1(x)$ denote the set of parameter values that satisfy \eqref{eq:case1_res1}-\eqref{eq:case1_res2}.

\bigskip
\noindent\textbf{Case 2} ($\lambda_1=0$, $\lambda_2>0$): By $\lambda_2>0$ and \eqref{eq:KKT3}, we obtain
$$\omega=q_{\theta_1}(1,0|x)=(1-\Phi_{2})\Phi_{1}+\Phi(x^{(1)}{}'\delta^{(1)}+\beta^{(1)})[\Phi_{2}-\Phi(x^{(2)}{}'\delta_{2}+\beta^{(2)})],$$
and $q_{\theta_1}(0, 1|x)=(1-\Phi(x^{(1)}{}'\delta^{(1)}+\beta^{(1)}))\Phi_{2}.$ Note that $\lambda_{2}>0$ iff
\begin{align*}
\frac{z_{1}}{z_1+\omega}<\frac{\Phi_{2}(1-\Phi_{1})}{z_2-\omega},
\end{align*}
which is equivalent to
\begin{align}(1-\Phi_{2})\Phi_{1}+\Phi(x^{(1)}{}'\delta^{(1)}+\beta^{(1)})[\Phi_{2}-\Phi(x^{(2)}{}'\delta_{2}+\beta^{(2)})]>\frac{z_{1}z_{2}-\Phi_{2}(1-\Phi_{1})z_{1}}{\Phi_{2}+\Phi_{1}-2\Phi_{1}\Phi_{2}}.\label{eq:case1_res3}
\end{align}
We let $\Theta_2(x)$ denote the set of parameter values that satisfy \eqref{eq:case1_res3}.

\bigskip
\noindent\textbf{Case 3} ($\lambda_1>0$, $\lambda_2=0$): By $\lambda_1>0$ and \eqref{eq:KKT2}, we obtain
$$\omega=q_{\theta_1}(1,0|x)=(1-\Phi(x^{(2)}{}'\delta_{2}+\beta^{(2)}))\Phi_{1},$$
and hence $q_{\theta_1}(0, 1|x)=(1-\Phi_{1})\Phi_{2}+\Phi(x^{(2)}{}'\delta_{2}+\beta^{(2)})[\Phi_{1}-\Phi(x^{(1)}{}'\delta_{1}+\beta^{(1)})].$ Note that $\lambda_{1}>0$ iff
$$\frac{z_{1}}{z_1+\omega}>\frac{\Phi_{2}(1-\Phi_{1})}{z_2-\omega},$$
which is equivalent to
\begin{align}(1-\Phi(x^{(2)}{}'\delta_{2}+\beta^{(2)}))\Phi_{1}<\frac{z_{1}z_{2}-\Phi_{2}(1-\Phi_{1})z_{1}}{\Phi_{2}+\Phi_{1}-2\Phi_{1}\Phi_{2}}.\label{eq:case1_res4}\end{align}
We let $\Theta_3(x)$ denote the set of parameter values that satisfy \eqref{eq:case1_res4}.

\subsubsection{Score Function}
We let $s_\theta=(s_{\beta^{(1)}},s_{\beta^{(2)}},s_{\delta^{(1)}},s_{\delta^{(2)}})'.$ Each component of $s_\theta$ takes the following form:
\begin{align}
    s_{\vartheta}(y|x)=\sum_{\bar y\in \mathcal Y}1\{y=\bar y\}z_{\vartheta}(\bar y|x),~\vartheta\in\{\beta^{(1)},\beta^{(2)},\delta^{(1)},\delta^{(2)}\},
\end{align}
where $z_\vartheta(\bar y|x)$ is the partial derivative of $\ln p_{\theta}(\bar y|x)$ with respect to $\vartheta$, which is well-defined if $\theta$ is in $\Theta_2(x)$, $\Theta_3(x)$, or in the interior of $\Theta_1(x)$. Let 
\begin{align}
   r_h(y,x)\equiv (\sqrt{q_{\theta+h}(y|x)}-\sqrt{q_\theta(y|x)}-\frac{1}{2}h's_\theta (y|x)\sqrt {q_\theta(y|x)})^2.
\end{align}
Suppose $\theta\in\Theta_2(x)$. By \eqref{eq:def_Theta2}, $\theta+h\in\Theta_2(x)$ for $\|h\|$ small enough. Then, pointwise,
$r_h(y,x)=o(\|h^2\|)$ because $s_\theta(y|x)=2\frac{1}{\sqrt{q_{\theta}(y|x)}}\frac{\partial}{\partial\theta}\sqrt{q_\theta(y|x)}=\frac{\partial}{\partial\theta}\ln q_\theta(y|x)$. The same argument applies when $\theta\in\Theta_3(x)$ or $\theta\in \text{int}(\Theta_1(x))$. The only case this argument does not apply is when $\theta$ is on the boundary between $\Theta_2(x)$ and $\Theta_1(x)$ (or between $\Theta_3(x)$ and $\Theta_1(x)$). For example, suppose $\theta$ is on the boundary between $\Theta_2(x)$ and $\Theta_1(x)$. Then, we may have $\theta+h\in\Theta_2(x)$ for all $h$ with $\|h\|>0$ but $\theta\in\Theta_1$. Then, the pointwise argument above does not apply. However, $\theta$ being on the boundary between the two sets means
\begin{align*}
    \frac{z_{1}z_{2}-\Phi_{2}(1-\Phi_{1})z_{1}}{\Phi_{2}+\Phi_{1}-2\Phi_{1}\Phi_{2}}=\Phi_{1}(1-\Phi_{2})+\Phi(x^{(1)}{}'\delta^{(1)}+\beta^{(1)})[\Phi_{2}-\Phi(x^{(2)}{}'\delta^{(2)}+\beta^{(2)})].
\end{align*}
If $x$ contains a continuous component (e.g. distance from headquarters/distribution center), the set of $x$'s satisfying above has measure 0, $r_h(y,x)$ is bounded on the set. Hence, it does not affect the integral in Assumption \ref{as:direc-diff}. Hence, Assumption \ref{as:direc-diff}
 holds. 

For completeness, the functional form of $z_\vartheta(\bar y|x)$ is derived below for each $\bar y\in\mathcal Y$ and $\vartheta\in\{\beta^{(1)},\beta^{(2)},\delta^{(1)},\delta^{(2)}\}.$
Across all subcases analyzed in the previous section, the form of $q_{\theta}(0, 0|x)$ and $q_{\theta}(1, 1|x)$ remains the same. We calculate score functions first by taking the pointwise derivative of $\ln q_{\theta}(0,0|x)$ and $\ln q_\theta(1,1|x)$. This yields 
\begin{align*}
&z_{\beta^{(1)}}(0, 0|x) = 0, \quad z_{\beta^{(2)}}(0, 0|x) = 0, \\
&z_{\delta^{(1)}}(0, 0|x) = -\frac{\phi_{1}x^{(1)}{}'}{(1-\Phi_{1})},\quad z_{\delta^{(2)}}(0, 0|x) = -\frac{\phi_{2}x^{(2)}{}'}{(1-\Phi_{2})}\\
&z_{\beta^{(1)}}(1,1|x) = \frac{\phi(x^{(1)}{}'\delta^{(1)}+\beta^{(1)})}{\Phi(x^{(1)}{}'\delta^{(1)}+\beta^{(1)})}, \quad z_{\beta^{(2)}}(1,1|x) = \frac{\phi(x^{(2)}{}'\delta^{(2)}+\beta^{(2)})}{\Phi(x^{(2)}{}'\delta^{(2)}+\beta^{(2)})}\\
&z_{\delta^{(1)}}(1,1|x) = \frac{\phi(x^{(1)}{}'\delta^{(1)}+\beta^{(1)})x^{(1)}}{\Phi(x^{(1)}{}'\delta^{(1)}+\beta^{(1)})}, \quad z_{\delta^{(2)}}(1,1|x) = \frac{\phi(x^{(2)}{}'\delta^{(2)}+\beta^{(2)})x^{(2)}}{\Phi(x^{(2)}{}'\delta^{(2)}+\beta^{(2)})},
\end{align*}
where $\phi_j=\phi(x^{(j)}{}'\delta^{(j)}),j=1,2$.

Next, we derive $z_{\vartheta}(1,0|x)$ and $z_\vartheta(0,1|x)$.

\noindent \textbf{Case 1:} Suppose $\theta\in\Theta_1(x)$. By taking the pointwise derivative of $\ln q_\theta$ in \eqref{eq:LF_density3}, one can obtain
\begin{align*}
z_{\delta^{(1)}}(1,0|x)&=\frac{\phi_{1}x^{(1)}}{\Phi_{1}}+\frac{\phi_{1}x^{(1)}-\phi_{1}\Phi_{2}x^{(1)}-\phi(x^{(1)}{}'\delta^{(1)}+\beta^{(1)})\Phi(x^{(2)}{}'\delta^{(2)}+\beta^{(2)})x^{(1)}}{\Phi_{1}+\Phi_{2}-\Phi_{1}\Phi_{2}-\Phi(x^{(1)}{}'\delta^{(1)}+\beta^{(1)})\Phi(x^{(2)}{}'\delta^{(2)}+\beta^{(2)})}-\frac{\phi_{1}x^{(1)}(1-2\Phi_{2})}{\Phi_{1}+\Phi_{2}-2\Phi_{1}\Phi_{2}}\\[1em]
z_{\delta^{(2)}}(1,0|x)&=\frac{-\phi_{2}x^{(2)}}{1-\Phi_{2}}+\frac{\phi_{2}x^{(2)}-\phi_{2}\Phi_{1}x^{(2)}-\phi(x^{(2)}{}'\delta^{(2)}+\beta^{(2)})\Phi(x^{(1)}{}'\delta^{(1)}+\beta^{(1)})x^{(2)}}{\Phi_{1}+\Phi_{2}-\Phi_{1}\Phi_{2}-\Phi(x^{(1)}{}'\delta^{(1)}+\beta^{(1)})\Phi(x^{(2)}{}'\delta^{(2)}+\beta^{(2)})}-\frac{\phi_{2}x^{(2)}(1-2\Phi_{1})}{\Phi_{1}+\Phi_{2}-2\Phi_{1}\Phi_{2}}\\[1em]
z_{\beta^{(1)}}(1,0|x)&=\frac{-\phi(x^{(1)}{}'\delta^{(1)}+\beta^{(1)})\Phi(x^{(2)}{}'\delta^{(2)}+\beta^{(2)})}{\Phi_{1}+\Phi_{2}-\Phi_{1}\Phi_{2}-\Phi(x^{(1)}{}'\delta^{(1)}+\beta^{(1)})\Phi(x^{(2)}{}'\delta^{(2)}+\beta^{(2)})}\\[1em]
z_{\beta^{(2)}}(1,0|x)&=\frac{-\phi(x^{(2)}{}'\delta^{(2)}+\beta^{(2)})\Phi(x^{(1)}{}'\delta^{(1)}+\beta^{(1)})}{\Phi_{1}+\Phi_{2}-\Phi_{1}\Phi_{2}-\Phi(x^{(1)}{}'\delta^{(1)}+\beta^{(1)})\Phi(x^{(2)}{}'\delta^{(2)}+\beta^{(2)})}.
\end{align*}
Similarly,
\begin{align*}
z_{\delta^{(1)}}(0,1|x)&=\frac{-\phi_{1}x^{(1)}}{1-\Phi_{1}}+\frac{\phi_{1}x^{(1)}-\phi_{1}\Phi_{2}x^{(1)}-\phi(x^{(1)}{}'\delta^{(1)}+\beta^{(1)})\Phi(x^{(2)}{}'\delta^{(2)}+\beta^{(2)})x^{(1)}}{\Phi_{1}+\Phi_{2}-\Phi_{1}\Phi_{2}-\Phi(x^{(1)}{}'\delta^{(1)}+\beta^{(1)})\Phi(x^{(2)}{}'\delta^{(2)}+\beta^{(2)})}-\frac{\phi_{1}x^{(1)}(1-2\Phi_{2})}{\Phi_{1}+\Phi_{2}-2\Phi_{1}\Phi_{2}}\\[1em]
z_{\delta^{(2)}}(0,1|x)&=\frac{\phi_{2}x^{(2)}}{\Phi_{2}}+\frac{\phi_{2}x^{(2)}-\phi_{2}\Phi_{1}x^{(2)}-\phi(x^{(2)}{}'\delta^{(2)}+\beta^{(2)})\Phi(x^{(1)}{}'\delta^{(1)}+\beta^{(1)})x^{(2)}}{\Phi_{1}+\Phi_{2}-\Phi_{1}\Phi_{2}-\Phi(x^{(1)}{}'\delta^{(1)}+\beta^{(1)})\Phi(x^{(2)}{}'\delta^{(2)}+\beta^{(2)})}-\frac{\phi_{2}x^{(2)}(1-2\Phi_{1})}{\Phi_{1}+\Phi_{2}-2\Phi_{1}\Phi_{2}}\\[1em]
z_{\beta^{(1)}}(0,1|x)&=\frac{-\phi(x^{(1)}{}'\delta^{(1)}+\beta^{(1)})\Phi(x^{(2)}{}'\delta^{(2)}+\beta^{(2)})}{\Phi_{1}+\Phi_{2}-\Phi_{1}\Phi_{2}-\Phi(x^{(1)}{}'\delta^{(1)}+\beta^{(1)})\Phi(x^{(2)}{}'\delta^{(2)}+\beta^{(2)})}\\[1em]
z_{\beta^{(2)}}(0,1|x)&=\frac{-\phi(x^{(2)}{}'\delta^{(2)}+\beta^{(2)})\Phi(x^{(1)}{}'\delta^{(1)}+\beta^{(1)})}{\Phi_{1}+\Phi_{2}-\Phi_{1}\Phi_{2}-\Phi(x^{(1)}{}'\delta^{(1)}+\beta^{(1)})\Phi(x^{(2)}{}'\delta^{(2)}+\beta^{(2)})}.
\end{align*}
\noindent\textbf{Case 2:} Suppose $\theta\in\Theta_2(x)$. Similarly to the analysis in Case 1, we may obtain
\begin{align*}
z_{\delta^{(1)}}(1,0|x)&=\frac{x^{(1)}(1-\Phi_{2})\phi_{1}+x^{(1)}\Phi_{2}\phi(x^{(1)}{}'\delta^{(1)}+\beta_{1})-x^{(1)}\Phi(x^{(2)}{}'\delta^{(2)}+\beta_{2})\phi(x^{(1)}{}'\delta^{(1)}+\beta_{1})}{(1-\Phi_{2})\Phi_{1}+\Phi(x^{(1)}{}'\delta^{(1)}+\beta^{(1)})[\Phi_{2}-\Phi(x^{(2)}{}'\delta^{(2)}+\beta^{(2)})]}\\[1em]
z_{\delta^{(2)}}(1,0|x)&=\frac{-x^{(2)}\Phi_{1}\phi_{2}+x^{(2)}\phi_{2}\Phi(x^{(1)}{}'\delta^{(1)}+\beta_{1})-x^{(2)}\Phi(x^{(1)}{}'\delta^{(1)}+\beta_{1})\phi(x^{(2)}{}'\delta^{(2)}+\beta_{2})}{(1-\Phi_{2})\Phi_{1}+\Phi(x^{(1)}{}'\delta^{(1)}+\beta^{(1)})[\Phi_{2}-\Phi(x^{(2)}{}'\delta^{(2)}+\beta^{(2)})]}\\[1em]
z_{\beta^{(1)}}(1,0|x)&=\frac{(\Phi_{2}-\Phi(x^{(2)}{}'\delta^{(2)}+\beta^{(2)}))\phi(x^{(1)}{}'\delta^{(1)}+\beta^{(1)})}{(1-\Phi_{2})\Phi_{1}+\Phi(x^{(1)}{}'\delta^{(1)}+\beta^{(1)})[\Phi_{2}-\Phi(x^{(2)}{}'\delta^{(2)}+\beta^{(2)})]}\\[1em]
z_{\beta^{(2)}}(1,0|x)&=\frac{\Phi(x^{(1)}{}'\delta^{(1)}+\beta^{(1)})\phi(x^{(2)}{}'\delta^{(2)}+\beta^{(2)})}{(1-\Phi_{2})\Phi_{1}+\Phi(x^{(1)}{}'\delta^{(1)}+\beta^{(1)})[\Phi_{2}-\Phi(x^{(2)}{}'\delta^{(2)}+\beta^{(2)})]},
\end{align*}
and
\begin{align*}
z_{\delta^{(1)}}(0,1|x)&=-\frac{x^{(1)}\phi(x^{(1)}{}'\delta^{(1)}+\beta^{(1)})}{1-\Phi(x^{(1)}{}'\delta^{(1)}+\beta^{(1)})},~~z_{\delta^{(2)}}(0,1|x)=\frac{x^{(2)}\phi_{2}}{\Phi_{2}}\\[1em]
z_{\beta^{(1)}}(0,1|x)&=-\frac{\phi(x^{(1)}{}'\delta^{(1)}+\beta^{(1)})}{1-\Phi(x^{(1)}{}'\delta^{(1)}+\beta^{(1)})},~~
z_{\beta^{(2)}}(0,1|x)=0.
\end{align*}
\noindent\textbf{Case 3:} Suppose $\theta\in\Theta_3(x)$. Similarly to the previous two cases, we may obtain
\begin{align*}
z_{\delta^{(1)}}(1,0|x)&=x^{(1)}\phi_{1}/\Phi_{1},~~z_{\delta^{(2)}}(1,0|x)=\frac{-x^{(2)}\phi(x^{(2)}{}'\delta^{(2)}+\beta^{(2)})}{1-\Phi(x^{(2)}{}'\delta^{(2)}+\beta^{(2)})}\\[1em]
z_{\beta^{(1)}}(1,0|x)&=0,~~
z_{\beta^{(2)}}(1,0|x)=\frac{-\phi(x^{(2)}{}'\delta^{(2)}+\beta^{(2)})}{1-\Phi(x^{(2)}{}'\delta^{(2)}+\beta^{(2)})},
\end{align*}
and
\begin{align*}
z_{\delta^{(1)}}(0,1|x)&=\frac{-x^{(1)}\Phi_{2}\phi_{1}+\Phi(x^{(2)}{}'+\delta^{(2)})x^{(1)}(\phi_{1}-\phi(x^{(1)}{}'\delta^{(1)}+\delta^{(1)}))}{(1-\Phi_{1})\Phi_{2}+\Phi(x^{(2)}{}'\delta^{(2)}+\beta^{(2)})[\Phi_{1}-\Phi(x^{(1)}{}'\delta^{(1)}+\beta^{(1)})]}\\[1em]
z_{\delta^{(2)}}(0,1|x)&=\frac{x^{(2)}(1-\Phi_{1})\phi_{2}+x^{(2)}(\Phi_{1}-\Phi(x^{(1)}{}'\delta_{1}+\beta_{1}))\phi(x^{(2)}{}'\delta_{2}+\beta_{2})}{(1-\Phi_{1})\Phi_{2}+\Phi(x^{(2)}{}'\delta^{(2)}+\beta^{(2)})[\Phi_{1}-\Phi(x^{(1)}{}'\delta^{(1)}+\beta^{(1)})]}\\[1em]
z_{\beta^{(1)}}(0,1|x)&=\frac{-\Phi(x^{(2)}{}'\delta^{(2)}+\beta^{(2)})\phi(x^{(1)}{}'\delta^{(1)}+\beta^{(1)})}{(1-\Phi_{1})\Phi_{2}+\Phi(x^{(2)}{}'\delta^{(2)}+\beta^{(2)})[\Phi_{1}-\Phi(x^{(1)}{}'\delta^{(1)}+\beta^{(1)})]}\\[1em]
z_{\beta^{(2)}}(0,1|x)&=\frac{(\Phi_{1}-\Phi(x^{(1)}{}'\delta^{(1)}+\beta^{(1)}))\phi(x^{(2)}{}'\delta^{(2)}+\beta^{(2)})}{(1-\Phi_{1})\Phi_{2}+\Phi(x^{(2)}{}'\delta^{(2)}+\beta^{(2)})[\Phi_{1}-\Phi(x^{(1)}{}'\delta^{(1)}+\beta^{(1)})]}.
\end{align*}

\subsection{Triangular Model with an Incomplete Control Function}
\subsubsection{Sharp Identifying Restrictions and Assumption \ref{as:hypotheses}}
Let us simplify $G$ in Example \ref{ex:incomplete_cf} first. Let $\mathcal U=\mathbb R.$ We assume $\beta> 0$ throughout, but a similar analysis can be done by assuming $\beta< 0.$ Suppose $d_i=1$ first. By \eqref{eq:outcome}-\eqref{eq:inc_cf1},
$y_i=0$ if 
\begin{align}
    u_i<-\alpha-w_i'\eta-\beta v_i,~\text{ for some }v_i\in[-z_i'\gamma,\infty).
\end{align}
 Then, by $\beta> 0$, we can write this event as $u_i\in \bigcup_{v\in [-z_i'\gamma,\infty)}(-\infty ,-\alpha-w_i'\eta-\beta v)= (-\infty,-\alpha-w_i'\eta+\beta z_i'\gamma)$.
By \eqref{eq:outcome}-\eqref{eq:inc_cf1} again, $y_i=1$ if 
\begin{align}
    u_i\ge -\alpha-w_i'\eta-\beta v_i,~\text{ for some }v_i\in[-z_i'\gamma,\infty).
\end{align}
This means that $y_i=1$ is consistent with the model whenever $u_i\in \bigcup_{v\in [-z_i'\gamma,\infty)}[-\alpha-w_i'\eta-\beta v,\infty)=\mathbb R$. These predictions can be summarized as
\begin{align}
    G(u_i|1,w_i,z_i;\theta)=\begin{cases}
    \{1\} & u_i\ge -\alpha-w_i'\eta+\beta z_i'\gamma\\
    \{0,1\} & u_i< -\alpha-w_i'\eta+\beta z_i'\gamma.
    \end{cases}\label{eq:G_cf1}
\end{align}

\begin{figure}[htbp]
\begin{center}
\begin{tikzpicture}[scale=0.8, domain=0:15,>=latex]
	\draw[->] (0,0) -- (0,7) node[left] {$u_i$};
	\draw[-] (0,0) -- (6,0);
	\draw[dashed] (0,2.4) -- (6,2.4) ;
	\draw (0,2.4) node[anchor=east] {$ -\alpha-w_i'\eta+\beta z_i'\gamma$};
	\draw (3,1.2) node {$ \{1\}$};
	\draw (3,4.5)   node { $\{0,1\}$};
	\draw (3,-.5) node {$d_i=0$};
	
	\draw[->] (9,0) -- (9,7) node[left] {$u_i$};
	\draw[-] (9,0) -- (15,0);
	\draw[dashed] (9,4.4) -- (15,4.4) ;
	\draw (9,4.4) node[anchor=east] {$-w_i'\eta+\beta z_i'\gamma$};
	\draw (12,2.5) node {$ \{0\}$};
	\draw (12,5.5)   node { $\{0,1\}$};
	\draw (12,-.5) node {$d_i=1$};	
\end{tikzpicture}
\caption{The set of predicted outcomes $G(u|d_i,w_i,z_i;\theta)$  when $\beta>0$}
\label{fig:inc_cf1}
\end{center}
\end{figure}
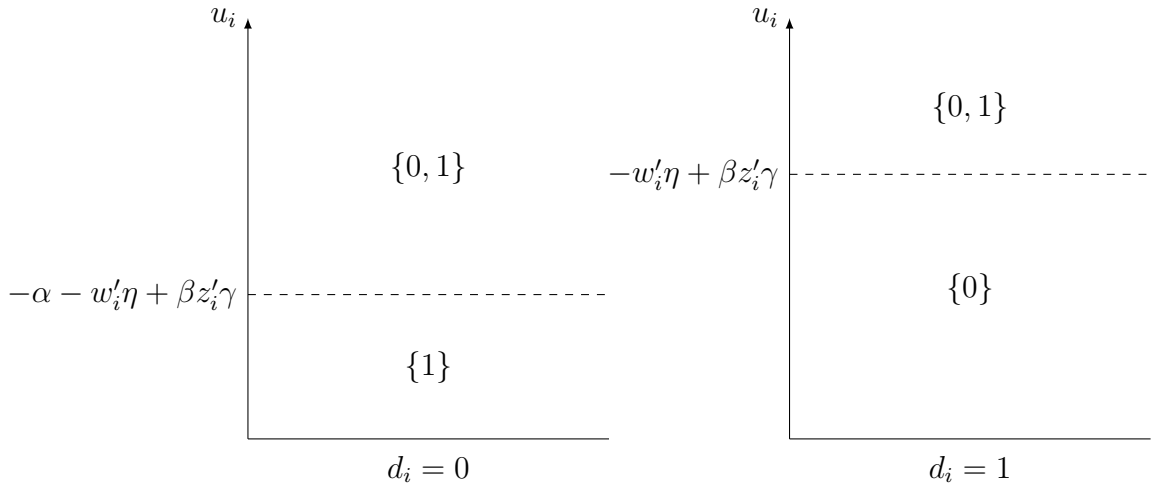

Now suppose $d_i=0$ implying $v_i\in (-\infty,-z_i'\gamma)$. Repeating a similar analysis yields the following correspondence
\begin{align}
     G(u_i|0,w_i,z_i;\theta)=\begin{cases}
    \{0\} & u_i\le -w_i'\eta+\beta z_i'\gamma\\
    \{0,1\} & u_i> -w_i'\eta+\beta z_i'\gamma.
    \end{cases}
\end{align}
These predictions are summarized in Figure \ref{fig:inc_cf1}. 

Assumption \ref{as:hypotheses} (i) holds because, as argued in Section \ref{sec:completeness}, the model's prediction under the null hypothesis is complete and is characterized by the reduced form function:
\begin{align}
    g(u_i|d_i,w_i,z_i;\theta)=1\{\alpha d_i+w_i'\eta+u_i\ge 0 \}.
\end{align}
This structure induces a unique conditional density for $y_i$. Suppose $u_i\sim N(0,1)$.\footnote{Here, we normalize the scale by setting the variance of $u_i$ to 1. Other choices of normalization are also possible.} 
Assumption \ref{as:hypotheses} (ii) holds as long as $z_i'\gamma<0$ with positive probability. For this, we demonstrate that there exists an event $A$ such that $\nu_{\theta_1}(A|x)>\nu^*_{\theta_0}(A|x)$ for some value of $x=(d,w,z)$. For this, take $A=\{1\}$ and suppose $d_i=0$. Under the null hypothesis, the conditional probability of $y_i=1$ is uniquely determined as $\nu_{\theta_0}^*(1|d_i=0,w_i,z_i)=\Phi(w_i'\eta)$. When $\beta>0$, \eqref{eq:G_cf1} implies
\begin{align}
    \nu_{\theta_1}(\{1\}|d_i=0,w_i,z_i)=\Phi(w_i'\eta-\beta z_i'\gamma),
\end{align}
which is greater than $\nu_{\theta_0}^*(1|d_i=0,w_i,z_i)$ for values of $z_i$ such that $z_i'\gamma <0$. Hence, $\mathfrak q_{\theta_0}$ and $\mathfrak q_{\theta_1}$ are disjoint.

\subsubsection{Least Favorable Parametric Model}
For any $\theta=(\beta,\delta)$ with $\beta>0$, the set $\mathfrak q_\theta$ of densities compatible with $\theta$ is then characterized by the following inequalities
\begin{align}
    q(0|d=0,w,z)&\ge \Phi(-w_i'\eta+\beta z_i'\gamma)\\
    q(1|d=0,w,z)&\ge 0,
\end{align}
and
\begin{align}
    q(0|d=1,w,z)&\ge 0\\
    q(1|d=1,w,z)&\ge 1-\Phi(-\alpha-w_i'\eta+\beta z_i'\gamma).
\end{align}

Suppose $d=0$. Let $z=q_1(0|d=0,w,z)$. Then, the convex program in \eqref{eq:cvx_prog}-\eqref{eq:cvx_prog2} can be written as
\begin{align}
    \min_{(q_0,q_1)}&\ln \Big(\frac{q_0(0|x)+z}{q_0(0|x)}\Big)(q_0(0|x)+z)+\ln\Big(\frac{1-q_0(0|x)+1-z}{1-q_0(0|x)}\Big)(q_0(1|x)+1-z)\\
    s.t.~ &q_0(y|x)=1-\Phi(w'\eta)\label{eq:ex4_const1}\\
    &z\ge \Phi(-w'\eta+\beta z'\gamma)\label{eq:ex4_const2}\\
    &1-z\ge 0,\label{eq:ex4_const3}
\end{align}
where \eqref{eq:ex4_const1} is due to the completeness of the model under the null hypothesis (see \eqref{eq:ex4_condprob_null}) and $d=0$.  Note that \eqref{eq:ex4_const3} is redundant since $q_1$ being in the probability simplex is implicitly assumed.
The KKT conditions associated with the program are, therefore
\begin{align}
    &\ln \frac{q_0(0|x)+z}{q_0(0|x)}-\ln\frac{2-q_0(0|x)-z}{1-q_0(0|x)}-\lambda=0\label{eq:ex4_kkt1}\\
    &\lambda(\Phi(-w'\eta+\beta z'\gamma)-z)\label{eq:ex4_kkt2}\\
    &\lambda\ge0\label{eq:ex4_kkt3}
\end{align}
where $q_0(y|x)=1-\Phi(w'\eta)$. There are two cases to consider. 

\bigskip
\noindent
Case 1 ($\lambda=0$): When $\lambda=0$, \eqref{eq:ex4_kkt1} implies $z=q_0(y|x)=1-\Phi(w'\eta)$. This holds when $z=1-\Phi(w'\eta)\ge \Phi(-w'\eta+\beta z'\gamma)$.

\bigskip
\noindent
Case 2 ($\lambda>0$): When $\lambda>0$, $z=\Phi(-w'\eta+\beta z'\gamma)$ by \eqref{eq:ex4_const2}. This occurs when
\begin{align}
    \lambda=\ln \frac{q_0(0|x)+z}{q_0(0|x)}-\ln\frac{2-q_0(0|x)-z}{1-q_0(0|x)}>0,
\end{align}
which is equivalent to
\begin{align}
    \Phi(-w'\eta+\beta z'\gamma)>q_0(y|x)=1-\Phi(w'\eta).
\end{align}

In sum, we have
\begin{align}
   q_{\theta_1}(0|d=0,w,z)=\begin{cases}
   1-\Phi(w'\eta)&\text{ if }\Phi(-w'\eta+\beta z'\gamma)\le 1-\Phi(w'\eta),\\
   \Phi(-w'\eta+\beta z'\gamma)&\text{ if }\Phi(-w'\eta+\beta z'\gamma)>1-\Phi(w'\eta),
   \end{cases} \label{eq:ex4_q0}
\end{align}
and $q_{\theta_1}(1|d=0,w,z)=1-q_{\theta_1}(0|d=0,w,z)$.
Repeating a similar analysis for $d=1$ yields
\begin{align}
   q_{\theta_1}(1|d=1,w,z)=\begin{cases}
   \Phi(\alpha+w'\eta)&\text{ if }1-\Phi(-\alpha-w'\eta+\beta z'\gamma)\le \Phi(\alpha +w'\eta),\\
  1-\Phi(-\alpha-w'\eta+\beta z'\gamma)&\text{ if }1-\Phi(-\alpha-w'\eta+\beta z'\gamma)> \Phi(\alpha +w'\eta),
   \end{cases} \label{eq:ex4_q1}
\end{align}
and $q_{\theta_1}(0|d=1,w,z)=1-q_{\theta_1}(1|d=1,w,z)$.
Recalling $\beta>0$ and $\Phi$ is strictly increasing, we may summarize \eqref{eq:ex4_q0}-\eqref{eq:ex4_q1} as follows
\begin{align}
    q_{\theta_1}(0|d=0,w,z)&=\begin{cases}
   \Phi(-w'\eta)&\text{ if } z'\gamma\le 0,\\
   \Phi(-w'\eta+\beta z'\gamma)&\text{ if }z'\gamma> 0,
   \end{cases} \\
       q_{\theta_1}(1|d=0,w,z)&=\begin{cases}
   1-\Phi(-w'\eta)&\text{ if } z'\gamma\le 0,\\
   1-\Phi(-w'\eta+\beta z'\gamma)&\text{ if }z'\gamma> 0,
   \end{cases} 
\end{align}
and
\begin{align}
    q_{\theta_1}(0|d=1,w,z)&=\begin{cases}
   \Phi(-\alpha-w'\eta)&\text{ if } z'\gamma\ge 0,\\
   \Phi(-\alpha-w'\eta+\beta z'\gamma)&\text{ if }z'\gamma< 0,
   \end{cases} \\
       q_{\theta_1}(1|d=1,w,z)&=\begin{cases}
   1-\Phi(-\alpha-w'\eta)&\text{ if } z'\gamma\ge 0,\\
   1-\Phi(-\alpha-w'\eta+\beta z'\gamma)&\text{ if }z'\gamma< 0.
   \end{cases} 
\end{align}
\subsubsection{Score Function}
The corresponding score function with respect to $\beta$ is
\begin{align}
    s_{\beta}(0|d=0,w,z)&=\begin{cases}
    0&\text{ if } z'\gamma\le 0,\\
   \frac{\phi(-w'\eta+\beta z'\gamma)}{\Phi(-w'\eta+\beta z'\gamma)}z'\gamma&\text{ if }z'\gamma> 0,
   \end{cases} \\
       s_{\beta}(1|d=0,w,z)&=\begin{cases}
   0&\text{ if } z'\gamma\le 0,\\
   -\frac{\phi(-w'\eta+\beta z'\gamma)}{\Phi(-w'\eta+\beta z'\gamma)}z'\gamma&\text{ if }z'\gamma> 0,
   \end{cases} 
\end{align}
and
\begin{align}
    s_{\beta}(0|d=1,w,z)&=\begin{cases}
   0&\text{ if } z'\gamma\ge 0,\\
   \frac{\phi(-\alpha-w'\eta+\beta z'\gamma)}{1-\Phi(-\alpha-w'\eta+\beta z'\gamma)}z'\gamma&\text{ if }z'\gamma< 0,
   \end{cases} \\
       s_{\beta}(1|d=1,w,z)&=\begin{cases}
   0&\text{ if } z'\gamma\ge 0,\\
   -\frac{\phi(-\alpha-w'\eta+\beta z'\gamma)}{1-\Phi(-\alpha-w'\eta+\beta z'\gamma)}z'\gamma&\text{ if }z'\gamma< 0.
   \end{cases} 
\end{align}

\subsection{Panel Dynamic Discrete Choice Models}\label{sec:panel_ddc}
 For each $t$, let $U_{it}=\alpha_i+\epsilon_{it}$.
We explicitly derive a form of $G$ below. Note that, $y_i=(y_{i1},y_{i2})=(0,0)$ occurs if 
\begin{align}
    u_{i1}<-x_{i1}'\eta,~u_{i2}<-x_{i2}'\eta, \label{eq:event1}
\end{align}
which follows from \eqref{eq:panel2}-\eqref{eq:panel3} 
or 
\begin{align}
    u_{i1}<-x_{i1}'\eta-\beta,~u_{i2}<-x_{i2}'\eta,\label{eq:event2}
\end{align}
which follows from  \eqref{eq:panel5}-\eqref{eq:panel6}. When $\beta\ge 0$, the union of the two events reduces to \eqref{eq:event1}.

Similarly, $y=(0,1)$ occurs if
\begin{align}
   u_{i1}<-x_{i1}'\eta,~u_{i2}\ge-x_{i2}'\eta, \label{eq:event3}
\end{align}
or 
\begin{align}
     u_{i1}<-x_{i1}'\eta-\beta,~u_{i2}\ge-x_{i2}'\eta.\label{eq:event4}
\end{align}
When $\beta\ge 0$, the union of the two events reduces \eqref{eq:event3}.

The outcome $y=(1,0)$ occurs if
\begin{align}
     u_{i1}\ge-x_{i1}'\eta,~u_{i2}<-x_{i2}'\eta-\beta, \label{eq:event5}
\end{align}
or 
\begin{align}
     u_{i1}\ge-x_{i1}'\eta-\beta,~u_{i2}<-x_{i2}'\eta-\beta, \label{eq:event6}
\end{align}
When $\beta\ge 0$, the union of the two events reduces to \eqref{eq:event6}.

The outcome $y=(1,1)$ occurs if
\begin{align}
     u_{i1}\ge-x_{i1}'\eta,~u_{i2}\ge-x_{i2}'\eta-\beta, \label{eq:event7}
\end{align}
or 
\begin{align}
     u_{i1}\ge-x_{i1}'\eta-\beta,~u_{i2}\ge-x_{i2}'\eta-\beta, \label{eq:event8}
\end{align}
When $\beta\ge 0$, the union of the two events reduces to \eqref{eq:event8}.
These predictions are summarized in Figure \ref{fig:ddc_levelset}.

The correspondence can therefore be written as
\begin{align}
    G(u_i|x_i;\theta)=\begin{cases}
    \{(0,0)\} &  u_{i1}<-x_{i1}'\eta-\beta,~u_{i2}<-x_{i2}'\eta,\\
    \{(0,1)\} &  u_{i1}<-x_{i1}'\eta-\beta,~u_{i2}\ge-x_{i2}'\eta,\\
    \{(1,0)\} &  u_{i1}\ge-x_{i1}'\eta,~u_{i2}<-x_{i2}'\eta-\beta,\\
    \{(1,1)\} &  u_{i1}\ge-x_{i1}'\eta,~u_{i2}\ge-x_{i2}'\eta-\beta,\\
    \{(0,0),(1,0)\} & -x_{i1}'\eta-\beta\le u_{i1}< -x_{i1}'\eta,~ u_{i2}\le -x_{i2}'\eta-\beta,\\
    \{(0,0),(1,1)\} & -x_{i1}'\eta-\beta\le u_{i1}< -x_{i1}'\eta,~  -x_{i2}'\eta-\beta\le u_{i2}<-x_{i2}'\eta,\\
    \{(0,1),(1,1)\} & -x_{i1}'\eta-\beta\le u_{i1}< -x_{i1}'\eta,~ u_{i2}\ge-x_{i2}'\eta.
    \end{cases}
\end{align}
A similar analysis can be done for the setting with $\beta\le 0$, which we omit for brevity.

\subsubsection{Sharp Identifying Restrictions and Assumption \ref{as:hypotheses}}
Assumption \ref{as:hypotheses} (i) holds because, as argued in \eqref{eq:ex3_unique}, the model makes a complete prediction with the following reduced-form function when $\beta=0$:
\begin{align}
    g(u_i|x_i;\theta)=\begin{bmatrix}
    1\{x_{i1}'\eta+\alpha_i+\epsilon_{i1}\ge 0\}\\
    1\{x_{i2}'\eta+\alpha_i+\epsilon_{i2}\ge 0\}
    \end{bmatrix}.
\end{align} Assumption \ref{as:hypotheses} (ii) holds if $U$ follows a distribution $F$ that is absolutely continuous with respect to the Lebesgue measure on $\mathbb R^2$. We show below, when $\beta>0$, there exists an event $A$ such that $\nu_{\theta_1}(A)>\nu_{\theta_0}^*(A)$ for all $\theta_1\in\Theta_1.$ For example, take $A=\{(1,1)\}$. As shown on the left panel of Figure \ref{fig:ddc_levelset}, the probability of $\{(1,1)\}$ is uniquely determined when $\beta=0$.
Therefore, the upper bound on the probability of $\{(1,1)\}$ is
\begin{align}
    \nu_{\theta_0}^*(\{(1,1)\}|x)=F(u_{i1}\ge-x_{i1}'\eta,~u_{i2}\ge-x_{i2}'\eta).
\end{align}
When $\beta>0$, the lower bound on the probability of the same event is
\begin{align}
    \nu_{\theta_1}(\{(1,1)\}|x)=F(u_{i1}\ge-x_{i1}'\eta,~u_{i2}\ge-x_{i2}'\eta-\beta),
\end{align}
which exceeds $\nu_{\theta_0}^*(\{(1,1)\}|x)$ as long as $F$ is absolutely continuous. This means $\mathfrak q_{\theta_1}$ and $\mathfrak q_{\theta_0}$ are disjoint.

\subsubsection{Least Favorable Parametric Model}
The analysis of the LFP and score is similar to that of discrete games. For brevity, we give the LFP below and omit its derivation.

For each $t\in \{1,2\}$, let 
$\Phi_{t}=\Phi(x'_{it}\eta+\gamma a)$ and $\Phi_{t\beta}= \Phi(x'_{it}\eta+\beta+\gamma a)$.
Define the following parameter sets.
\begin{align}
\Theta_1(x)&=\Big\{\theta\in\Theta: \frac{\int\Phi_{1}\Phi_{2\beta}\phi(a)da}{\int\Phi_{1}\Phi_{2}\phi(a)da}\geq \frac{\int\Phi_{1\beta}(1-\Phi_{2\beta})\phi(a)da}{\int \Phi_{1}(1-\Phi_{2})\phi(a)da}\Big\}\\
\Theta_2(x)&=\Big\{\theta\in\Theta:\frac{\int\Phi_{1}\Phi_{2\beta}\phi(a)da}{\int\Phi_{1}\Phi_{2}\phi(a)da}\geq \frac{1-\int[(1-\Phi_{1})\Phi_{2}+\Phi_{1}\Phi_{2\beta}]\phi(a)da}{\int (1-\Phi_{2})\phi(a)da},\notag\\
&\qquad\frac{\int\Phi_{1}(1-\Phi_{2})\phi(a)da}{\int(1-\Phi_{1})(1-\Phi_{2})\phi(a)da}\leq \min\Big\{\frac{\nu^{*}_\theta(\{(1,0)\}|x)}{K_\theta-\nu^{*}_\theta(\{(1,0)\}|x)}, \frac{K_\theta-\nu_\theta(\{(0,0)\}|x)}{\nu_\theta(\{(0,0)\}|x)}\Big\}\Big\},
\end{align}
where
\begin{align}
K_\theta&=1-\int (1-\Phi_{1})\Phi_{2}\phi(a)da-\int\Phi_{1}\Phi_{2\beta}\phi(a)da\\
\nu^*_\theta(\{(1,0)\}|x)&=\int \Phi_{1\beta}(1-\Phi_{2\beta})\phi(a)da\\
\nu_\theta(\{(0,0)\}|x)&=\int (1-\Phi_{1\beta})(1-\Phi_2)\phi(a)da.	
\end{align}
When $\theta\in\Theta_1(x)$,  the least favorable parametric model is characterized by the following density:
\begin{align}
q_{\theta}((0,0)|x) 
& = \int (1-\Phi_{1})(1-\Phi_{2})\phi(a)da - \int (\Phi_{1}-\Phi_{1\beta})(1-\Phi_{2\beta})\phi(a)da\\
q_{\theta}((0,1)|x) &= \int (1-\Phi_{1})\Phi_{2}\phi(a)da, \\
q_{\theta}((1,0)|x) &=\int \Phi_{1\beta}(1-\Phi_{2\beta})\phi(a)da, \\
q_{\theta}((1,1)|x) &= \int\Phi_{1}\Phi_{2\beta}\phi(a)da. 
\label{eqn: LFP nonmixing}    
\end{align}
When $\theta\in\Theta_2(x)$, the least favorable parametric model is characterized by the following density:
\begin{align}
q_{\theta}((0,0)|x) &=\frac{\int(1-\Phi_1)(1-\Phi_2)\phi(a)da}{\int(1-\Phi_2)\phi(a)da}K_\theta, \\
q_{\theta}((0,1)|x) &= \int (1-\Phi_{1})\Phi_{2}\phi(a)da, \\
q_{\theta}((1,0)|x) &=\frac{\int\Phi_1(1-\Phi_2)\phi(a)da}{\int(1-\Phi_2)\phi(a)da}K_\theta, \\
q_{\theta}((1,1)|x) &= \int\Phi_{1}\Phi_{2\beta}\phi(a)da.    
\end{align}

\section{Lemmas and Proofs}\label{sec:appdx_lemmas_proofs}
This section is organized as follows. Section \ref{ssec:lfmodel} contains the proof of Proposition \ref{prop:lfmodel}.
In Section \ref{ssec:deltahat}, we show $\sqrt n$-consistency of $\hat\delta_n$ by extending standard arguments for extremum estimators to locally incomplete models. In Section  \ref{ssec:size_control}, we use the results in \ref{ssec:deltahat} to show results on the asymptotic size of our score test.

\subsection{Least favorable parametric model}\label{ssec:lfmodel}

\begin{proof}[\rm Proof of Proposition \ref{prop:lfmodel}]
By Assumption \ref{as:hypotheses} (i), the model makes a complete prediction for any $\theta\in\Theta_0.$ Then, for each $\theta\in\Theta$, $q_\theta$ is the unique density in $\mathfrak q_\theta$ satisfying $q_\theta=dQ_\theta/d\mu$, where $Q_\theta(\cdot|x)=F_\theta(g(U|x;\theta)\in \cdot).$ Hence, the LF parametric model is well-defined for $\theta\in\Theta_0.$ Now, let $\theta\in\tilde\Theta\setminus\Theta_0$. By Assumption \ref{as:hypotheses} (ii), $\mathfrak q_{\theta_0}\cap \mathfrak q_{\theta_h}\ne \emptyset$ for any $h\in (B_1-\beta_0)\cap \mathbb C_\epsilon$. Also, 
\begin{align}
    \mathfrak q_\theta=\Big\{q:q=\frac{dQ}{d\mu},~core(\nu_\theta)\Big\},
\end{align}
where $core(\nu_\theta)$ is the set of probability measures on $\mathcal Y$ such that $Q(A)\ge \nu_\theta(A)$ for all $A\subseteq \mathcal Y$. Note that $\nu_\theta$ is a belief function. Hence, there exists a least favorable pair $(Q_0,Q_h)\in core(\nu_{\theta_0})\times core(\nu_{\theta_h})$ by Theorem 3.1 in \cite{kz} applied with $n=1$. We then let $q_{\theta_h}=dQ_h/d\mu$. Recall that $\theta_h=(\beta+h,\delta)$. Hence, we also defined $q_\theta$ for $\theta\in\tilde\Theta\setminus\Theta_0$. This completes the proof.
\end{proof}

\subsection{$\sqrt n$-consistency of $\hat\delta_n$}\label{ssec:deltahat}

\begin{lemma}\label{lem:choquet_lip}
 Suppose Assumption \ref{as:ftheta_Lip} holds. Then for any bounded function $g:\mathcal Y\times\mathcal X\to \mathbb R$,
 \begin{align}
    \big|\int gd\nu^*_\theta-\int gd\nu^*_{\theta'}\big|\le C'\|\theta-\theta'\|,~\forall \theta,\theta'\in\Theta,\label{eq:lip2}
\end{align}
and
\begin{align}
    \big|\int gd\nu_\theta-\int gd\nu_{\theta'}\big|\le C'\|\theta-\theta'\|,~\forall \theta,\theta'\in\Theta,\label{eq:lip1}
\end{align}
for some $C'>0$.
\end{lemma}

\begin{proof}
Note that
\begin{align}
    \int gd\nu^*_\theta&=\int \max_{(y,x)\in G(u|x;\theta)\times \{x\}}g(y,x)dF_\theta(u)\notag\\
    &=\int \bar g(u)f_\theta(u)du,
\end{align}
where $\bar g(u)=\max_{(y,x)\in G(u|x;\theta)\times \{x\}}g(y,x)$. This, boundedness of $g$, and Assumption \ref{as:ftheta_Lip} imply
\begin{align}
\Big| \int gd\nu^*_\theta- \int gd\nu^*_{\theta'}\Big|& =\Big|\int \bar g(u)(f_\theta(u)-f_{\theta'}(u))du\Big|\notag\\
&\lesssim  \|f_\theta-f_{\theta'}\|_{L^1_\zeta} \lesssim \|\theta-\theta'\|.
\end{align}
This ensures \eqref{eq:lip2}. Showing \eqref{eq:lip1} is analogous and is omitted.
\end{proof}

The following proposition shows the sample log-likelihood converges to its population counterpart uniformly over a set of distributions consistent with the null or local alternative hypotheses. Recall that the set of conditional laws $\mathcal Q^n_\theta$ is defined as in \eqref{eq:defQn}, and $\mathcal P_{\theta}^n$ collects joint laws  of $(Y^n,X^n)$.
\begin{proposition}[ULLN]\label{prop:ulln}
	Suppose Assumptions \ref{as:hypotheses}-\ref{as:consistency} hold. Let $h\in\mathcal V_1.$
	Then, for any $\theta_0\in\Theta_0$ and $\epsilon>0$, there exists $N_\epsilon\in\mathbb N$ that does not depend on $\delta$ such that
\begin{align}
	\sup_{P^n\in\mathcal P^n_{\theta_0+h/\sqrt n}}P^n\Big(\sup_{\delta\in\Theta_\delta}\big|n^{-1}\sum_{i=1}^n\ln \q_{\beta_0,\delta}(Y_i|X_i)-E_{Q_0}[\ln \q_{\beta_0,\delta}]\big|\ge \epsilon \Big)<\epsilon,~\forall n\ge N_\epsilon.\label{eq:ulln}
\end{align}
\end{proposition}

\begin{proof}
Below, let $\nu_{\theta}$ and $\nu^*_{\theta}$ be a belief function and its conjugate induced by the correspondence $(u,x)\mapsto G(u|x;\theta)\times \{x\}$ on $\mathcal Y\times\mathcal X$. That is, they are set functions such that
\begin{align}
    \nu_{\theta}(A)&=\int_{\mathcal X}\int_{\mathcal U}1\{G(u|x;\theta))\times \{x\}\subseteq A\}dF_\theta(u)dq_x(x),~A\subset\mathcal Y\times \mathcal X\label{eq:def_nutheta}\\
    \nu_{\theta}^*(A)&=\int_{\mathcal X}\int_{\mathcal U}1\{G(u|x;\theta))\times \{x\}\cap A\ne \emptyset\}dF_\theta(u)dq_x(x).~A\subset\mathcal Y\times \mathcal X.\label{eq:def_nuthetastar}
\end{align}
A key observation is that, for any $\theta_0\in\Theta_0$,
\begin{align}
\int \ln \q_{\beta_0,\delta}d\nu_{\theta_0}	=E_{Q_{\theta_0}}[\ln \q_{\beta_0,\delta}]=\int \ln \q_{\beta_0,\delta}(y|x)d\nu^*_{\theta_0}.\label{eq:null_equality}
\end{align}
This is because the model is complete under $H_0$ by Assumption \ref{as:hypotheses} and the fact that the Choquet integrals with respect to $\nu_{\theta_0}$ and $\nu_{\theta_0}^*$ coincide with each other in such a setting.

Note that one may write the event (i.e., the argument of $P^n$) in \eqref{eq:ulln} as the union of the following two events:
\begin{align}
A^U_n&=\Big\{(y^n,x^n):\sup_{\delta\in\Theta_\delta}\Big(n^{-1}\sum_{i=1}^n\ln \q_{\beta_0,\delta}(y_i|x_i)-\int \ln \q_{\beta_0,\delta}d\nu^*_{\theta_0}\Big)\ge\epsilon\Big\}	\\
A^L_n&=\Big\{(y^n,x^n):\inf_{\delta\in\Theta_\delta}\Big(n^{-1}\sum_{i=1}^n\ln \q_{\beta_0,\delta}(y_i|x_i)-\int \ln \q_{\beta_0,\delta}d\nu_{\theta_0}\Big)\le -\epsilon\Big\}.
\end{align}
Let $K^n=\prod_{i=1}^n K_i$ be a random set whose distribution follows the law induced by $F_{\theta_0+h/\sqrt n}$. Below, we simply write $K^n\sim F_{\theta_0+h/\sqrt n}.$ Note that
\begin{align}
	\sup_{P^n\in\mathcal P^n_{\theta_0+h/\sqrt n}}P^n(A^U_n\cap A^L_n)&=F_{\theta_0+h/\sqrt n}\Big(K^n\cap (A^U_n\cup A^L_n)\ne\emptyset\Big)\\
	&\le F_{\theta_0+h/\sqrt n}\big(K^n\cap A^U_n\ne\emptyset\big)+F_{\theta_0+h/\sqrt n}\big(K^n\cap A^L_n\ne\emptyset\big)\\
	&=\nu^*_{\theta_0+h/\sqrt n}\Big(\sup_{\delta\in\Theta_\delta}\big[n^{-1}\sum_{i=1}^n\ln \q_{\beta_0,\delta}(Y_i|X_i)-\int \ln \q_{\beta_0,\delta}d\nu^*_{\theta_0}\big]\ge\epsilon\Big)\label{eq:concentration1}\\
	&\quad+\nu^*_{\theta_0+h/\sqrt n}\Big(\inf_{\delta\in\Theta_\delta}\big[n^{-1}\sum_{i=1}^n\ln \q_{\beta_0,\delta}(Y_i|X_i)-\int \ln \q_{\beta_0,\delta}d\nu_{\theta_0}\big]\le -\epsilon\Big).\label{eq:concentration2}
\end{align}
By Assumption \ref{as:consistency}  and Lemma \ref{lem:choquet_lip},
\begin{align}
  \Big| \int \ln \q_{\beta_0,\delta}d\nu^*_{\theta_0+h/\sqrt n}-\int \ln \q_{\beta_0,\delta}d\nu^*_{\theta_0}\Big|&\le \frac{C'|h|}{\sqrt n},
\end{align}
implying there exists $\bar\eta>0$ and $N_{\bar\eta}$ such that $\sqrt n\sup_{\delta\in\Theta_\delta}\Big(\int \ln \q_{\beta_0,\delta}d\nu^*_{\theta_0+h/\sqrt n}-\int \ln \q_{\beta_0,\delta}d\nu^*_{\theta_0}\Big)<\bar\eta$ for all $n\ge N_{\bar\eta}$. Hence, for all $n\ge N_{\bar\eta}$, \eqref{eq:concentration1} is bounded by
\begin{align}
	\nu^*_{\theta_0+h/\sqrt n}\Big(\sup_{\delta\in\Theta_\delta}\frac{1}{\sqrt n}\sum_{i=1}^n\big[\ln \q_{\beta_0,\delta}(Y_i|X_i)-\int \ln \q_{\beta_0,\delta}d\nu^*_{\theta_0+h/\sqrt n}\big]\ge\sqrt n\epsilon-\bar\eta\Big).\label{eq:concentration_centered1}
\end{align}
As shown below, we may apply Lemma  \ref{lem:tail_bounds} to this quantity.
Similarly, by Assumption \ref{as:consistency} and Lemma \ref{lem:choquet_lip}, \eqref{eq:concentration2} is bounded by
\begin{align}
	\nu^*_{\theta_0+h/\sqrt n}\Big(\inf_{\delta\in\Theta_\delta}\frac{1}{\sqrt n}\sum_{i=1}^n\big[\ln \q_{\beta_0,\delta}(Y_i|X_i)-\int \ln \q_{\beta_0,\delta}d\nu_{\theta_0+h/\sqrt n}\big]\le -\sqrt n\epsilon+\bar\eta\Big).\label{eq:concentration_centered2}
\end{align}

Now, let $\mathcal G\equiv\{g=\ln q_{\beta_0,\delta},\delta\in\Theta_\delta\}$. Then, by Lemma \ref{lem:bdd_lipschitz}, the induced family of functions $\mathcal F_{\mathcal G}$ defined in \eqref{eq:induced_class} consists of uniformly bounded and Lipschitz functions. By Theorem 2.7.11 in \cite{VanderVaarta}, it follows that
\begin{align}
	N_{[]}(\epsilon \|F_{\mathcal G}\|_{L^2(M)},\mathcal F_{\mathcal G},L^2(M))\le N(\epsilon/2, \Theta_\delta,\|\cdot\|)\le (2\text{diam}(\Theta_\delta)/\epsilon)^{d_{\delta}}.
\end{align}
Therefore, $\mathcal F_{\mathcal G}$ satisfies the condition of Lemma \ref{lem:tail_bounds}. Applying the lemma ensures that \eqref{eq:concentration_centered1} is bounded by
\begin{align}
	\Big(C_{\text{diam}(\Theta_\delta)}\frac{\sqrt n\epsilon-\bar \eta}{\sqrt d_\delta}\Big)^{d_\delta}e^{-2(\sqrt n\epsilon-\bar \eta)^2},
\end{align}
which tends to 0 as $n\to \infty$. \eqref{eq:concentration_centered2} can be handled similarly. This completes the proof.
\end{proof}

Let $S$ be a Euclidean space.
Given a family $\mathcal G$ of measurable functions on $S$  and a random closed set $K:\Omega\mapsto \mathcal K(S)$, define a family of measurable functions on $ \mathcal K(S)$ by
\begin{align}
	\mathcal F_{\mathcal G}\equiv\Big\{f:f(K)=\max_{s\in K}g(s),~g\in\mathcal G\Big\}.\label{eq:induced_class}
\end{align}
We denote the envelope function of $\mathcal F_{\mathcal G}$ by $F_{\mathcal G}$.
A class $\mathcal F$  of uniformly bounded functions is covered by at most $(\frac{D}{\epsilon})^{v}$ brackets  if for positive constants $v$ and $D$, 
\begin{align}
	 N_{[]}(\epsilon\|F\|_{L^2(M)} ,\mathcal F,L^2(M))\le \Big(\frac{D}{\epsilon}\Big)^{v},~~0<\epsilon<D,\label{eq:bracketing}
\end{align}

The following lemma gives concentration inequalities for the suprema (and infima) of empirical processes under plausibility functions.
\begin{lemma}\label{lem:tail_bounds}
	Let $\nu^n$ be a belief function such that $\nu^n(B)=M^n(K^n\subset A)$ for any $A\in \mathcal K(S^n)$.
	Let $\mathcal G$ be a family of uniformly bounded measurable functions on $S$ such that $\mathcal F_{\mathcal G}$ in \eqref{eq:induced_class} is covered by at most $(\frac{D}{\epsilon})^{v}$ brackets.  Then, for all $t>0$
\begin{align}
	&\nu^{*,n}\Big(\sup_{g\in\mathcal G}\frac{1}{\sqrt n}\sum_{i=1}^n\big[g(s_i)-\int gd\nu^*\big]\ge t\Big)\le \Big(C_D\frac{t}{\sqrt v}\Big)^ve^{-2t^2},\\
	&\nu^{*,n}\Big(\inf_{g\in\mathcal G}\frac{1}{\sqrt n}\sum_{i=1}^n\big[g(s_i)-\int gd\nu\big]\le -t\Big)\le \Big(C_D\frac{t}{\sqrt v}\Big)^ve^{-2t^2}.
\end{align}  
for some $C_D$ that depends on $D$ only.
\end{lemma}
\begin{proof}
Define the following events
\begin{align}
B^U_n&=\Big\{s^n:\sup_{g\in\mathcal G}\frac{1}{\sqrt n}\sum_{i=1}^n\big[g(s_i)-\int gd\nu^*\big]\ge t\Big\}	\\
B^L_n&=\Big\{s^n:\inf_{g\in\mathcal G}\frac{1}{\sqrt n}\sum_{i=1}^n\big[g(s_i)-\int gd\nu\big]\le - t\Big\}.
\end{align}
Observe that
\begin{align}
	K^n\cap B^U_n\ne \emptyset&~\Leftrightarrow~\sup_{s^n\in K^n}\sup_{g\in\mathcal G}\frac{1}{\sqrt n}\sum_{i=1}^n\big[g(s_i)-\int gd\nu^*\big]\ge t\notag\\
	&~\Leftrightarrow~\sup_{g\in\mathcal G}\frac{1}{\sqrt n}\sum_{i=1}^n\big[\sup_{s_i\in K_i}g(s_i)-\int gd\nu^*\big]\ge t\notag\\
	&~\Leftrightarrow~\sup_{g\in\mathcal G}\frac{1}{\sqrt n}\sum_{i=1}^n\big[\sup_{s_i\in K_i}g(s_i)-\int \sup_{s\in K}g(s)dM(K)\big]\ge t\notag\\
	&~\Leftrightarrow~\sup_{f\in\mathcal F_{\mathcal G}}\frac{1}{\sqrt n}\sum_{i=1}^n\big[f(K_i)-E_{M}[f(K)]\big]\ge t. \label{eq:ulln_max1}
\end{align}
Therefore,
\begin{align}
	\nu^{*,n}\Big(\sup_{g\in\mathcal G}\frac{1}{\sqrt n}\sum_{i=1}^n\big[g(s_i)-\int gd\nu^*\big]\ge  t\Big)&=M^n(K^n\cap B^U_n\ne\emptyset)\\
	&=M^n\Big(\sup_{f\in\mathcal F_{\mathcal G}}\frac{1}{\sqrt n}\sum_{i=1}^n\big[f(K_i)-E_{M}[f(K)]\big]\ge t\Big),
\end{align}
By Theorem 1.3 (ii) in \cite{Talagrand:1994aa}, for all $t>0$,
\begin{align}
	M^n\Big(\sup_{f\in\mathcal F_{\mathcal G}}\big(\frac{1}{\sqrt n}\sum_{i=1}^nf(K_i)-E_{M}[f(K)]\big)\ge t\Big)\le M^n\Big(\|\mathbb G_nf\|_{\mathcal F}\ge t\Big)\le \Big(C_D\frac{t}{\sqrt v}\Big)^ve^{-2t^2}.
\end{align}
A similar argument can be applied to $B^L_n$ as well.
\end{proof}

Let $K$ be a subset of $\mathcal S=\mathcal Y\times \mathcal X.$
The following lemma shows that $f_\delta(K)\equiv\max_{(y,x)\in K}\ln q_{\beta_0,\delta}(y|x)$ is uniformly bounded and Lipschitz, which provides a control of the covering number.
\begin{lemma}\label{lem:bdd_lipschitz}
	Suppose Assumption \ref{as:consistency} holds. Then, (i) $f_\delta$ is uniformly bounded; and (ii) for any $\delta,\delta'\in\Theta_\delta$,
	\begin{align}
	|f_\delta(K)-f_{\delta'}(K)|\lesssim\|\delta-\delta'\|.\label{eq:fdelta_lip}
	\end{align}
\end{lemma}
\begin{proof} 
(i) follows from the map $\delta\mapsto M(\delta)$ being continuous by Assumption \ref{as:consistency} and hence achieving a finite maximum on the compact set $\Theta_\delta$. 

(ii)
Let $s=(y,x)$ and let $g(\delta,s)=\ln q_{\beta_0,\delta}(y|x)$.
By Assumption \ref{as:consistency},
\begin{align}
    f_{\delta'}(K)=\max_{(y,x)\in K}\Big(g(\delta',s)-g(\delta,s)+g(\delta,s)\Big)\le \max_{s\in K}\Big(\|\delta-\delta'\|+g(\delta,s)\Big)=f_\delta(K)+\|\delta-\delta'\|.
\end{align}
Similarly,
\begin{align}
    f_{\delta'}(K)=\max_{s\in K}\Big(g(\delta',s)-g(\delta,s)+g(\delta,s)\Big)\ge \max_{s\in K}\Big(-\|\delta-\delta'\|+g(\delta,s)\Big)=f_\delta(K)-\|\delta-\delta'\|.
\end{align}
Combining the two inequalities above yields \eqref{eq:fdelta_lip}.
\end{proof}

\bigskip
Below, we write $X_n=O_{P^n}(a_n)$ uniformly in $P^n\in\mathcal F_n$ if for any $\epsilon>0$, there exist finite $M>0$ and $N>0$ such that $\sup_{P^n\in\mathcal F_n}P^n(|X_n/a_n|> M)<\epsilon$ for all $n> N$. 
\begin{theorem}\label{thm:consistency}
	Suppose Assumptions \ref{as:hypotheses}-\ref{as:consistency} hold. Then, for any $\eta>0$,
\begin{align}
\lim_{n\to\infty}		\inf_{P^n\in \mathcal P^n_{\theta_0+h/\sqrt n}}P^n\Big(\|\hat\delta_n-\delta_0\|< \eta\Big)=1
	\end{align}	
	and $\Mn(\hat\delta_n)\ge \Mn(\delta_0)-O_{P^n}(r_n^{-2})$ uniformly in $P^n\in\mathcal P^n_{\theta_0+h/\sqrt n} $.
\end{theorem}
\begin{proof}
The proof is based on the standard argument for the consistency of extremum estimators \citep[see, e.g.][]{newey1994large}. A slight difference is that one needs a uniform law of large numbers under any sequence $P^n\in\mathcal P^n_{\theta_0+h/\sqrt n}$, which is established by Proposition \ref{prop:ulln}.
For each $\delta\in\Theta_\delta$, recall that $\M(\delta)\equiv E_{Q_0}[\ln \q_{\beta_0,\delta}]$ and let $\Mn(\delta)\equiv n^{-1}\sum_{i=1}^n\ln \q_{\beta_0,\delta}(s_i).$ 
Given any neighborhood $V$ of $\delta_0$, we want to show that $\hat \delta_n\in V$, wp$\to 1$ uniformly over $\mathcal P^n_{\theta_0+h/\sqrt n}$. For this, it suffices to show that $\inf_{P^n\in \mathcal P^n_{\theta_0+h/\sqrt n}}P^n(\M(\hat\delta_n)<\inf_{\delta\in\Theta\cap V^c}\M(\delta))\to 1$. 
Let $\epsilon\equiv \inf_{\delta\in\Theta\cap V^c}\M(\delta)- \M(\delta_0)$. This constant is well-defined since $\inf_{\Theta\cap V^c}\M(\delta)=\M(\delta^*)>\M(\delta_0)$ for some $\delta^*\in \Theta\cap V^c$ by Assumption \ref{as:consistency} and the compactness of $\Theta_\delta$.  

Let $A_{1n}\equiv\{\omega:\M(\hat \delta_n) <\Mn(\hat\delta_n)+\epsilon/3 \}, A_{2n}\equiv\{\omega:\Mn(\hat\delta_n)<\Mn(\delta_0)+\epsilon/3\},A_{3n}\equiv\{\omega:\Mn(\delta_0)<\M(\delta_0)+\epsilon/3\}$. 
For any $\omega\in A_{1n}\cap A_{2n}\cap A_{3n}$, 
\begin{align*}
   \M(\hat \delta_n) &<\Mn(\hat\delta_n)+\epsilon/3  \\
&<\Mn(\delta_0)+2\epsilon/3\\
&<\M(\delta_0)+\epsilon.
\end{align*}           
Therefore, 
\begin{align}
	\inf_{P^n\in\mathcal P^n_{\theta_0+h/\sqrt n}}P^n(\M(\hat \delta_n)<\M(\delta_0)+\epsilon)&\ge \inf_{P^n\in\mathcal P^n_{\theta_0+h/\sqrt n}} P^n( A_{1n}\cap A_{2n}\cap A_{3n})\\
	&\ge\inf_{P^n\in\mathcal P^n_{\theta_0+h/\sqrt n}} \Big(1-P^n(A^c_{1n})-P(A^c_{2n})-P(A^c_{3n})\Big)\\
	&\ge 1-\sum_{j=1}^3\sup_{P^n\in\mathcal P^n_{\theta_0+h/\sqrt n}}P^n(A^c_{jn}).
\end{align} 
Note that, for any $h$, $\sup_{P^n\in\mathcal P^n_{\theta_0+h/\sqrt n}}P^n(A_{1n}^c)\to 0$ and $\sup_{P^n\in\mathcal P^n_{\theta_0+h/\sqrt n}}P^n(A_{3n}^c)\to 0$ by Proposition \ref{prop:ulln}.  Also note that $\sup_{P^n\in\mathcal P^n_{\theta_0+h/\sqrt n}} P^n(A_{2n}^c)\to 0$ by the construction of $\hat\delta_n$.
\end{proof}

\begin{proof}[\rm Proof of Proposition \ref{prop:rootn-consistency}]
We use results from \cite{VanderVaarta} (Section 3.2.2) to establish the desired result. 
By Assumption \ref{as:majorant}, for every $\delta$ in a neighborhood of $\delta_0$ 
\begin{align}
    E_P[\ln q_{\beta_0,\delta}-\ln q_{\beta_0,\delta_0}] \lesssim -\|\delta-\delta_0\|^2.
\end{align}
Now let $\mathbb G_n f=\frac{1}{\sqrt n}\sum_{i=1}^n f(Y_i,X_i)-E[f(Y_i,X_i)]$ be an empirical process  defined on $\mathcal M_\zeta=\{\ln q_{\beta_0,\delta}-\ln q_{\beta_0,\delta_0}:\|\delta-\delta_0\|\le \zeta\}$ for $\zeta>0$. By Assumption \ref{as:consistency} (i-b) and arguing as in Example 3.2.12 in \cite{Van-der-Vaart:2000aa}, we obtain
 \begin{align}
     E^*_{P^n}[\|\mathbb G_n\|_{\mathcal M_\zeta}]\le \zeta.
 \end{align}
This bound holds uniformly over $P^n\in \mathcal P_{\theta}$ for any $\theta$ in a neighborhood of $\theta_0$.
Note also that, by Assumption \ref{as:consistency} (iii), $\hat\delta_n$ satisfies $\Mn(\hat\delta_n)\ge \inf_{\delta\in\Theta_\delta}\Mn(\delta)+r_n$, for some sequence  $\{r_n\}$, and for any $\epsilon>0$ and $\theta$ in a neighborhood of $\theta_0$, $\sup_{P^n\in\mathcal P^n_{\theta}}P^n(|r_n|>\epsilon)\to 0$.
 The result then follows from Corollary 3.2.6 in \cite{VanderVaarta} with $\phi_n(\zeta)=\zeta$, $r_n=\sqrt n$, and applying their argument for $P^n\in \mathcal P^n_{\theta_0+h/\sqrt n}.$
\end{proof}

\subsection{Asymptotic Size of the Score Test}\label{ssec:size_control}

\begin{proof}[\rm Proof of Theorem \ref{thm:size}]

 To simplify the exposition, we assume $V_0=E_{P_0}[s_\beta(Y_i|X_i;\beta_0,\delta_0)s_\beta(Y_i|X_i;\beta_0,\delta_0)']$ is known for now. Let
\begin{align}
	g^*_n(\beta_0)=\frac{1}{\sqrt n}\sum_{i=1}^n s_\beta(Y_i|X_i;\beta_0,\delta_0).
\end{align}
 By Assumption \ref{as:hypotheses}, $P_0^n=Q_{\theta_0}^n\times q_X^n$ for some unique product measure.
By Assumption \ref{as:direc-diff} and arguing as in Theorem 7.2 in \cite{Van-der-Vaart:2000aa}, we have $E[s_\beta(Y_i|X_i;\beta_0,\delta_0)]=E[s_\delta(Y_i|X_i;\beta_0,\delta_0)]=0$, where expectation is with respect to $P_0$.
By the square integrability of $s_\beta$ ensured by Assumption \ref{as:direc-diff}, the central limit theorem for i.i.d. sequences ensures
\begin{align}
g^*_n(\beta_0)    \stackrel{P^n_{0}}{\leadsto}N(0,V_0),\label{eq:size1}
\end{align}
Define
\begin{align}
S^*_n=	n g^*_n(\beta_0)'V_0^{-1} g^*_n(\beta_0)-\inf_{h\in \mathcal V_1}n( g^*_n(\beta_0)-h)' V_0^{-1}( g^*_n(\beta_0)-h).
\end{align}
By \eqref{eq:size1} and the continuous mapping theorem, it then follows that 
\begin{align}
    S^*_n\stackrel{P^n_{0}}{\leadsto} S,\label{eq:Sstar_conv}
\end{align}
where $S$ is as in \eqref{eq:defS}. 

For the desired result, it remains to show that $S_n$ is asymptotically equivalent to $S^*_n$ under $P^n_{0}$. 
For each $\delta,$ let $\mathbb G_ns_\beta(\delta)\equiv\frac{1}{n}\sum_{i=1}^n s_\beta(Y_i|X_i;\beta_0,\delta)-E[s_\beta(Y_i|X_i;\beta_0,\delta)].$
We may then write
\begin{align}
 g_n(\beta_0)-g^*_n(\beta_0)&=\mathbb G_n s_\beta(\hat\delta_n)-\mathbb G_n s_\beta(\delta_0)\notag\\
&\quad -\sqrt n(E[s_\beta(Y_i|X_i;\beta_0,\hat\delta_n)]-E[s_\beta(Y_i|X_i;\beta_0,\delta_0)])\notag\\
&=o_{P^n}(1)+\frac{\partial}{\partial \delta}E[s_\beta(Y_i|X_i;\beta_0,\delta)]\big|_{\delta=\tilde\delta_n}\sqrt n(\hat\delta_n-\delta_0)
=o_{P^n}(1),\label{eq:g-gstar}
\end{align}
where the second equality follows from the stochastic equicontinuity of $\mathbb G_ns_\beta$ and applying the mean-value theorem componentwise, where $\tilde\delta_n$ may be different across different components.
The last equality follows from the $\sqrt n$-consistency of $\hat\delta_n$ ensured by Proposition \ref{prop:rootn-consistency}  and 
\begin{align}
  \frac{\partial}{\partial \delta}E[s_\beta(Y_i|X_i;\beta_0,\delta)]\big|_{\delta=\delta_0}=0.\label{eq:ortho}
\end{align}
The equality above holds because the score equation is the first-order condition for a maximization problem, and $\hat\delta_n$ has a limit that maximizes the same objective function \citep[see][p.1357-1358]{newey1994asymptotic}. Consider maximizing $E[\ln q_{\beta,\delta}(y|x)]$ with respect to $\delta$. The first-order conditions for this maximization are $\partial E[\ln q_{\beta,\delta}(y|x)]/\partial\delta=0$ identically in $\beta$. Taking a derivative with respect to $\beta$ and evaluating it at $(\beta_0,\delta_0) $ yields \eqref{eq:ortho}.

Let $\varphi(x)=x'V_0^{-1}x-\inf_{h\in\mathcal V_1}(x-h)'V_0^{-1}(x-h)$. Note that $x\mapsto \inf_{h\in\mathcal V_1}(x-h)'V_0^{-1}(x-h)$ is continuous due to Berge's maximum theorem \citep[][Theorem 17.31]{aliprantisborder}. Hence, $\varphi$ is continuous. By \eqref{eq:g-gstar} and the continuous mapping theorem,
\begin{align}
\hat S_n-S^*_n=\varphi(\sqrt n g_n(\beta_0))-\varphi(\sqrt n g^*_n(\beta_0))=o_{P^n}(1).\label{eq:S-Sstar}
\end{align}
By \eqref{eq:Sstar_conv} and \eqref{eq:S-Sstar},
\begin{align}
	\lim_{n\to\infty}P^n_{0}(\hat S_n>c_\alpha)=\alpha.
\end{align}
This establishes the claim of the theorem. 

Note that we assumed $I_{\theta_0}$ was known. In general, it can be consistently estimated by $\hat I_n=n^{-1}\sum_{i=1}^n s_\theta(Y_i|X_i;\beta_0,\hat\delta_n)s_\theta(Y_i|X_i;\beta_0,\hat\delta_n)'$. To see this, let $s_{\theta,j}$ be the $j$-th component of $s_\theta$. For each $j$ and $k$, define
\begin{align}
\xi_{j,k}(Y_i,X_i;\delta)\equiv s_{\theta,j}(Y_i|X_i;\beta_0,\delta)s_{\theta,k}(Y_i|X_i;\beta_0,\delta).
\end{align}
For the  $(j,k)$-th component of $\hat I_n$, we then have
\begin{align*}
    [\hat I_n]_{j,k}-[I_{\theta_0}]_{j,k}&=\frac{1}{n}\sum_{i=1}^n\xi_{j,k}(Y_i,X_i;\hat\delta_n) -E_{q_{\theta_0}}[\xi_{j,k}(Y_i,X_i;\delta_0)]\\
    &=\big(\frac{1}{n}\sum_{i=1}^n\xi_{j,k}(Y_i,X_i;\hat\delta_n)-E_{q_{\theta_0}}[\xi_{j,k}(Y_i,X_i;\hat\delta_n)\big)\\
    &\qquad+\big(E_{q_{\theta_0}}[\xi_{j,k}(Y_i,X_i;\hat\delta_n)-E_{q_{\theta_0}}[\xi_{j,k}(Y_i,X_i;\delta_0)\big)=o_{P^n}(1),
\end{align*}
where the last equality follows because of $\sup_{\delta}|\frac{1}{n}\sum_{i=1}^n\xi_{j,k}(Y_i,X_i;\delta)-E_{q_{\theta_0}}[\xi_{j,k}(Y_i,X_i;\delta)|=o_{P^n}(1)$ due to Assumption \ref{as:gc}, $\hat\delta_n$'s consistency by Theorem \ref{thm:consistency}, and the continuity of $\delta\mapsto E[\xi_{j,k}(Y_i,X_i;\delta)]$. Given this, showing the claim of the theorem with the estimated $I_{\theta_0}$ is straightforward by applying Slutsky's theorem.
\end{proof}

\clearpage
\section{Figures}
\begin{figure}[htbp]
    \centering
    \includegraphics[trim=0 200 0 200, clip, scale=.75]{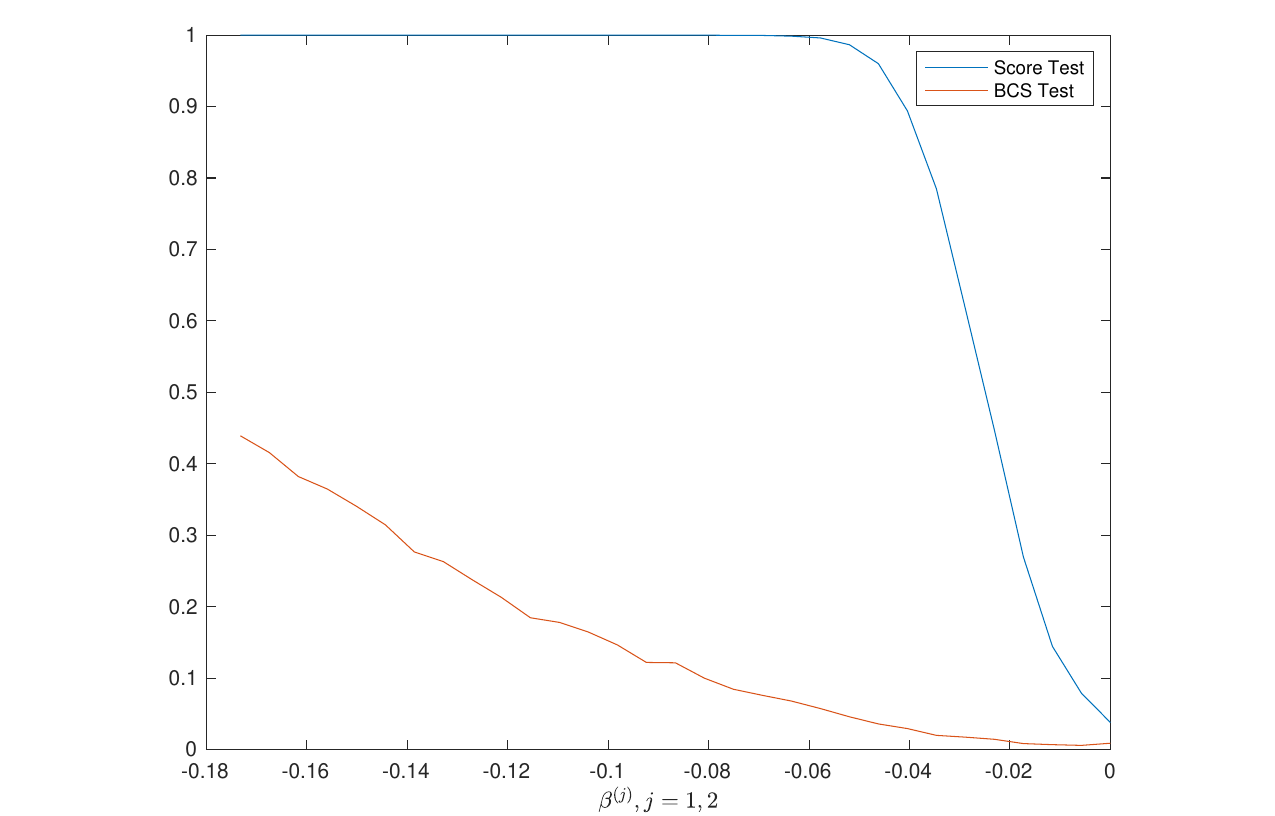}
    \caption{Power of the Score and BCS Tests (Design 1)}
    \label{fig:power_iid}
\end{figure}
\begin{figure}[htbp]
    \includegraphics[trim=0 200 0 200, clip,scale=.75]{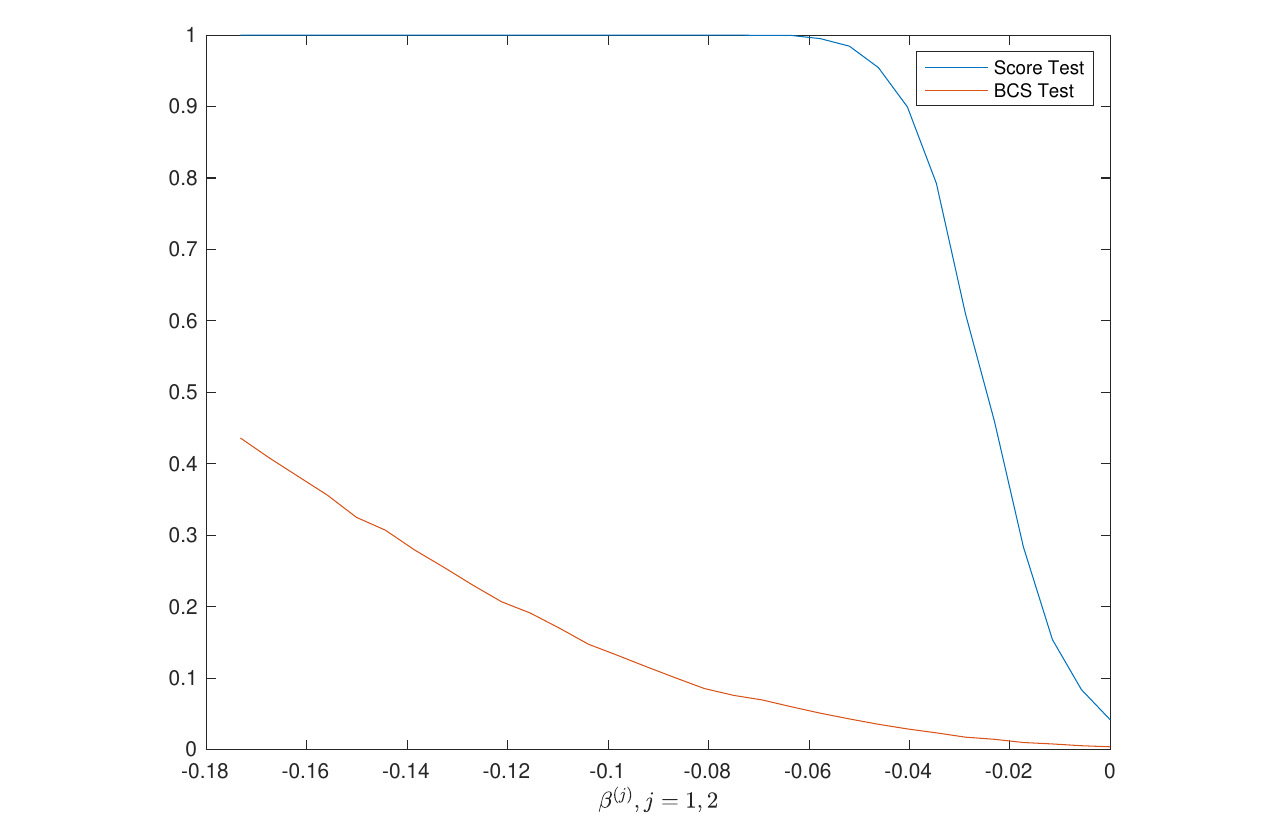}
    \caption{Power of the Score and BCS Tests (Design 2)}
    \label{fig:power_LFP}
\end{figure}
\end{document}